\documentclass[journal]{IEEEtran}
\usepackage[utf8]{inputenc}
\usepackage{times}
\usepackage{amsmath}
\usepackage{amssymb}
\usepackage{amsthm}
\usepackage{mathrsfs}
\usepackage{subcaption}
\usepackage{graphicx}
\usepackage{enumerate}
\usepackage[compress]{cite}
\usepackage{float}

\usepackage[acronym]{glossaries}
\usepackage{enumitem}
\usepackage{multirow}

\newacronym{3GPP}{3GPP}{third generation partnership project}
\newacronym{5G}{5G}{fifth generation}
\newacronym{6G}{6G}{sixth generation}
\newacronym{AP}{AP}{access point}
\newacronym{ASO}{ASO}{access point switch-on/off}
\newacronym[longplural={angles of arrival}]{AoA}{AoA}{angle of arrival}
\newacronym[longplural={angles of departure}]{AoD}{AoD}{angle of departure}
\newacronym{B5G}{B5G}{beyond 5G}
\newacronym{BCD}{BCD}{block coordinate descend}
\newacronym{BRPA}{BRPA}{balanced random pilot assignment}
\newacronym{BS}{BS}{base station}
\newacronym{CAP}{CAP}{compress-after-precoding}
\newacronym{CB}{CB}{conjugate beamforming}
\newacronym{CBP}{CBP}{compress-before-precoding}
\newacronym{CCDF}{CCDF}{complementary cumulative distribution function}
\newacronym{CDF}{CDF}{cumulative distribution function}
\newacronym{CF}{CF}{cell-free}
\newacronym{CF-mMIMO}{CF-mMIMO}{cell-free massive MIMO}
\newacronym{CMS}{CMS}{conventional mobile station}
\newacronym{CoMP}{CoMP}{coordinated multipoint}
\newacronym{CPA}{CPA}{cluster-based pilot assignment}
\newacronym{CPU}{CPU}{central processing unit}
\newacronym{C-RAN}{C-RAN}{cloud radio access network}
\newacronym{CSI}{CSI}{channel state information}
\newacronym{DCPA}{DCPA}{dissimilarity cluster-based pilot assignment}
\newacronym{DL}{DL}{downlink}
\newacronym{EMS}{EMS}{energy harvesting mobile station}
\newacronym{GOPA}{GOPA}{globally optimal power allocation}
\newacronym{IoT}{IoT}{Internet-of-Things}
\newacronym{LMMSE}{LMMSE}{linear minimum mean square error}
\newacronym{LS}{LS}{least-squares}
\newacronym{LSFD}{LSFD}{large-scale fading decoding}
\newacronym{LOS}{LOS}{line-of-sight}
\newacronym{MIMO}{MIMO}{multiple-input multiple-output}
\newacronym{MF}{MF}{matched filtering}
\newacronym{M-MIMO}{M-MIMO}{massive MIMO}
\newacronym{MMSE}{MMSE}{minimum mean square error}
\newacronym{mmWave}{mmWave}{millimeter wave}
\newacronym{MRC}{MRC}{maximal ratio combining}
\newacronym{MS}{MS}{mobile station}
\newacronym{MSE}{MSE}{mean square error}
\newacronym{MU-MIMO}{MU-MIMO}{multiuser-MIMO}
\newacronym{NCB}{NCB}{normalized conjugate beamforming}
\newacronym{NMRC}{NMRC}{normalized maximal ratio combiner}
\newacronym{NMSE}{NMSE}{normalized mean square error}
\newacronym{NOPA}{NOPA}{normalized optimal power allocation}
\newacronym{NLOS}{NLOS}{non-line-of-sight}
\newacronym{QoS}{QoS}{quality of service}
\newacronym{rms}{rms}{root mean square}
\newacronym{RHS}{RHS}{right hand side}
\newacronym{RPA}{RPA}{random pilot assignment}
\newacronym{RRM}{RRM}{radio resource management}
\newacronym{RF}{RF}{radio frequency}
\newacronym{SINR}{SINR}{signal-to-interference-plus-noise ratio}
\newacronym{SISO}{SISO}{single-input single-output}
\newacronym{SNR}{SNR}{signal-to-noise ratio}
\newacronym{SOC}{SOC}{second order cone}
\newacronym{SWIPT}{SWIPT}{simultaneous wireless information and power transfer}
\newacronym{TDD}{TDD}{time division duplexing}
\newacronym{UDN}{UDN}{ultra dense network}
\newacronym{ULA}{ULA}{uniform linear array}
\newacronym{UPA}{UPA}{uniform planar array}
\newacronym{UL}{UL}{uplink}
\newacronym{WPT}{WPT}{wireless power transfer}
\newacronym{ZF}{ZF}{zero-forcing}

\newcommand{\bs}{\boldsymbol}
\DeclareMathOperator{\SINR}{SINR}
\DeclareMathOperator{\diag}{diag}
\DeclareMathOperator{\blockdiag}{blockdiag}

\DeclareMathOperator{\Var}{Var}

\begin{document}

\title{SWIPT-enhanced Cell-Free Massive MIMO Networks}

\author{
Guillem~Femenias,~\IEEEmembership{Senior Member,~IEEE,}
Jan~Garc\'ia-Morales,~\IEEEmembership{Member,~IEEE,}
and
Felip~Riera-Palou,~\IEEEmembership{Senior Member,~IEEE}
\thanks{
G. Femenias, J. Garc\'ia-Morales and F. Riera-Palou are with the Mobile Communications Group, University of the Balearic Islands, Palma 07122, Illes Balears, Spain (e-mail: \{guillem.femenias,jan.garcia,felip.riera\}@uib.es).
}
\thanks{
This work was supported by the Ministerio de Ciencia, Innovaci\'on y Universidades (MCIU), the Agencia Estatal de Investigaci\'on (AEI), and the European Regional Development Fund (ERDF) through project TERESA (TEC2017-90093-C3-3-R).
}
\thanks{
Author Accepted Manuscript of the article:
G. Femenias, J. Garc\'ia-Morales, and F. Riera-Palou,
``SWIPT-enhanced Cell-Free Massive MIMO Networks,''
\emph{IEEE Transactions on Communications},
vol. 69, no. 8, pp. 5593--5607, Aug.~2021.
The final published version is available at:
https://doi.org/10.1109/TCOMM.2021.3083644.
Copyright \copyright\ 2021 IEEE.
}
}

% \markboth{IEEE Transactions on Communications}{Femenias \textit{et al.}: SWIPT-enhanced Cell-Free Massive MIMO Networks}

\maketitle

\begin{abstract}
Simultaneous wireless information and power transfer (SWIPT) has been advocated as a highly promising technology to provide near-\textit{perpetual} operation to low-powered wireless devices in Internet-of-Things (IoT)-based wireless networks. In this paper, a SWIPT-enhanced cell-free massive MIMO network is proposed. In such a network, a large set of spatially distributed access points (APs) interconnected via a central processing unit (CPU) can collaboratively serve a large number of both energy harvesting mobile stations (MSs) (requiring wireless energy transfer) and conventional MSs (not requiring wireless energy transfer) on the same time-frequency resources. We consider spatially correlated Rician fading channels and the use of different precoding schemes that are based on different channel estimators differing on the assumed knowledge of the line-of-sight component. Mathematically manageable expressions are derived for the harvested energy during the downlink (DL) energy harvesting phase and the \textit{achievable} spectral and energy efficiencies during the uplink (UL) payload transmission phase. A coupled UL/DL optimization problem is formulated aiming at finding the power control coefficients that maximize the minimum of the weighted achievable UL signal-to-interference-plus-noise ratios (SINRs) of all MSs. Extensive numerical results are presented that serve to highlight the existing trade-offs among the achievable spectral and energy efficiencies, the harvested energy, the energy dedicated to UL pilot transmission or the system configuration.
\end{abstract}


% \IEEEpeerreviewmaketitle

\section{Introduction}

\IEEEPARstart{N}{owadays}, the Internet has become ubiquitous and is transforming our lifestyle in unimaginable ways. But the journey still goes on, and we are now entering the era of the \gls{IoT} \cite{Al-Fuqaha15}. An era in which a massive number of intelligent devices and appliances will be connected to the network to share information and coordinate decisions so that the concepts of smart cities, smart homes, smart healthcare, or smart transportation, to name a few, become a reality \cite{Zanella14,Xu14,Islam15,Guerrero-Ibanez15,Lin17}. To achieve this formidable goal, several research challenges, including reliability, interoperability, security, and energy consumption will have to be addressed. However, according to recent research work on this topic, probably the most important challenge to overcome will be the bottleneck caused by energy management problems related to the short battery life of most wireless active components in the \gls{IoT} network \cite{Chu18}.

Under certain conditions, energy harvesting seems to be the most sensible option to provide near-\textit{perpetual} operation to such a massive amount of low-powered wireless devices in \gls{IoT} \cite{Kamalinejad15,Guo18}. Even though energy can be harvested from environmental sources, including, among others, solar, wind, thermal, or mechanical vibration, they are all uncertain and hence, \gls{WPT} is regarded as a more reliable technology to tackle the limited energy storage issue of \gls{IoT} devices \cite{LaRosa19}. Among the different incarnations of \gls{WPT} technology, \gls{SWIPT} seems to be a viable candidate for \gls{IoT} as it enables an acceptable level of certainty without an unaffordable increase in infrastructural requirements \cite{Chae18,Huang18}. One of the main challenges to be overcome by \gls{SWIPT}, however, are the large propagation path losses experienced in wireless scenarios, which will be even more exacerbated when considering \gls{mmWave} frequency bands. Even though the use of massive \gls{MIMO} can ameliorate the efficiency of \gls{SWIPT} in cellular-based wireless networks, the energy harvesting opportunities of cell-edge users will still be compromised \cite{Dong17}.

In trying to overcome some of the drawbacks of traditional cell-based massive \gls{MIMO} with co-located antennas, Ngo \textit{et al.} proposed a new paradigm, named \gls{CF-mMIMO} \cite{Ngo17}. Comprising of a large set of spatially distributed \glspl{AP} interconnected via a \gls{CPU}, \gls{CF-mMIMO} networks aim at collaboratively serving multiple \glspl{MS} on the same resources taken from a time-frequency grid. In this way, it is highly probable that, irrespective of their particular location in the coverage area of the network, each \gls{MS} is surrounded by a significant number of \glspl{AP}, thus experiencing a high degree of macro-diversity and a reduced level of propagation path losses. Macro-diversity helps in reducing the probability of blockage, and low propagation path loss levels are necessary to provide the required energy harvesting opportunities in \gls{SWIPT}-enhanced networks. The synergetic deployment of \gls{CF-mMIMO} and \gls{SWIPT} has been recently advocated by different authors in, for instance, \cite{Shrestha18,Alageli19,Demir20maxmin,Demir20LSFD,Wang20WPT,Wang20scheduling,Kusaladharma20}. In \cite{Shrestha18}, Shrestha and Amarasuriya consider the implementation of a \gls{SWIPT}-assisted \gls{CF-mMIMO} network where different transmission frames are scheduled for so-called energy and information users. Energy users harvest the energy by exploiting a \gls{DL} time switching protocol and then use the harvested energy for \gls{UL} transmission. \Gls{CB} and \gls{MF} are employed for \gls{DL} and \gls{UL} transmission, respectively, over Rayleigh fading channels. In \cite{Alageli19}, Alageli \textit{et al.} consider a cell-free massive
\gls{MIMO} where the \glspl{AP} serve a large number of information users and a single information-untrusted dual-antenna active energy-harvesting user that uses one antenna to legitimately harvest energy and the other antenna to eavesdrop information. Demir and Bj\"ornson in \cite{Demir20maxmin,Demir20LSFD} consider the implementation of power control algorithms to maximize the minimum \gls{UL} spectral efficiency for \gls{DL} \gls{WPT}-assisted \gls{CF-mMIMO}. They consider spatially uncorrelated Rician fading and maximum ratio processing with \gls{LSFD} based on different flavors of linear channel estimation. In \cite{Wang20WPT}, Wang \textit{et al.} propose a wirelessly powered cell-free \gls{IoT} system where the \gls{UL} and \gls{DL} power control coefficients are jointly optimized to minimize the total energy consumption assuming the use of distributed precoders and decoders. The implementation of long-term scheduling and power control allocation in wirelessly powered cell-free \gls{IoT} networks is investigated by the same authors in \cite{Wang20scheduling}. The coexistence and interplay between \gls{SWIPT} and \gls{CF-mMIMO} is characterized by Kusaladharma \textit{et al.} in \cite{Kusaladharma20} from a stochastic geometry-based perspective. Different from previous research works, which are based on a distributed cell-free operation over spatially uncorrelated channels, in this paper we propose a coupled \gls{UL}/\gls{DL} weighted max-min per-user rate optimization assuming the use of centralized cell-free operation and the transmission over spatially correlated Rician fading channels, a propagation environment that has been shown to be far more realistic \cite{Bjornson19}. Taking everything into account, our main contributions in this paper are as follows:
\begin{itemize}[noitemsep,wide=0pt, leftmargin=\dimexpr\labelwidth + 2\labelsep\relax]

   \item A \gls{SWIPT}-enhanced framework is proposed for a cell-free massive \gls{MIMO} scheme capable of servicing both energy harvesting \glspl{MS} (requiring wireless energy transfer) and conventional \glspl{MS} (not requiring wireless energy transfer). Mathematically manageable expressions are derived for the harvested energy at the \glspl{MS} and the \gls{UL} achievable spectral and energy efficiencies. Unlike prior art on this subject and in line with practical scenarios, \glspl{MS} can be either in \gls{LOS} or in \gls{NLOS} with respect to each of the \glspl{AP} in the network and the spatial correlation of the antenna arrays at the \glspl{AP} is also accounted for. Two channel estimators are considered that rely on different degrees of channel state information knowledge and that serve to upper- and lower-bound the performance that could be achieved in practical systems. Finally, the framework incorporates an estimation of the power consumed by the different components of the network (i.e., \glspl{AP}, \glspl{MS} and fronthaul links) thus allowing the derivation of realistic achievable energy efficiencies.

   \item A coupled \gls{UL}/\gls{DL} optimization problem is posed and solved that aims at finding the power control coefficients used by the energy harvesting \glspl{MS} during the \gls{DL} energy harvesting phase and the conventional \glspl{MS} during the \gls{UL} payload transmission stage. This optimization problem aims at maximizing the minimum (Max-Min) of the weighted achievable \gls{UL} \glspl{SINR} of all \glspl{MS} while satisfying the corresponding \gls{DL} and \gls{UL} transmit power constraints at both the \glspl{AP} and \glspl{MS}. Notably, the weights in the optimization process allow the balancing of the very different \gls{SINR} requirements conventional and energy-harvesting users potentially have.

   \item Extensive numerical results are presented that serve to highlight the existing trade-offs that might be needed among the amount of energy harvested during the \gls{DL} energy harvesting phase, the complexity of the Max-Min optimizations, the amount of energy dedicated to \gls{UL} pilot transmission or the system configuration (i.e., \gls{DL} energy harvesting phase length, \gls{UL} training phase length, number of active energy harvesting \glspl{MS} or spatial density of \glspl{AP}) and the achievable spectral and energy efficiencies.

\end{itemize}

This paper is organized as follows. The proposed \gls{SWIPT}-enhanced \gls{CF-mMIMO} system is described in Section \ref{sec:System_model}. Different subsections are devoted to present the general structure of the network and associated transmission protocols, the channel model, the large- and small-scale training phases, the \gls{DL} energy harvesting phase, and the \gls{UL} payload transmission phase. In Section \ref{sec:UL_DL_max-min}, a coupled \gls{UL}/\gls{DL} \textit{achievable} \gls{SINR} optimization problem is posed assuming the use of \gls{ZF} precoders/decoders. Section \ref{sec:Performance_metrics} presents the performance metrics (i.e., UL spectral and energy efficiencies) used in this paper. A discussion of the numerical results is provided in Section \ref{sec:numerical_results} and, finally, concluding remarks are recapped in Section \ref{sec:Conclusion}.

\textit{Notation}: Lower- and upper-case boldface symbols are used to denote vectors and matrices, respectively. Symbol $\boldsymbol{I}_q$ denotes a $q$-dimensional identity matrix. Given a matrix $\bs{X}$, the mathematical operators $\bs{X}^T$, $\bs{X}^*$, $\bs{X}^H$, and $\bs{X}^{-1}$ denote its transpose, conjugate, conjugate transpose, and inverse, respectively. The operator $\|\bs{x}\|$ represents the $l^2$-norm of vector $\bs{x}$, $[\bs{x}]_n$ is used to denote the $n$th element of vector $\bs{x}$, $\diag(\bs{x})$ denotes a diagonal matrix with the entries of vector $\bs{x}$ on its main diagonal, and $\blockdiag\left(\bs{X}_1 \ldots \bs{X}_n\right)$ denotes a block diagonal matrix comprising matrices $\bs{X}_1$, $\ldots$, $\bs{X}_n$ on its main block diagonal. The expectation and variance operators are represented by $\mathbb{E}\{\cdot\}$ and $\Var\{\cdot\}$, respectively. Concluding this notation paragraph, $\mathcal{N}(0,\sigma^2)$ is used to denote a zero-mean real valued Gaussian random variable with standard deviation $\sigma$, $\mathcal{CN}(\bs{m},\bs{R})$ will serve to denote a complex-valued Gaussian vector distribution with mean $\bs{m}$ and covariance $\bs{R}$, and $\mathcal{U}[a,b]$ represents a uniform random variable defined in the range $[a,b]$.

\begin{figure}
  \centering
  \includegraphics[width=.96\linewidth]{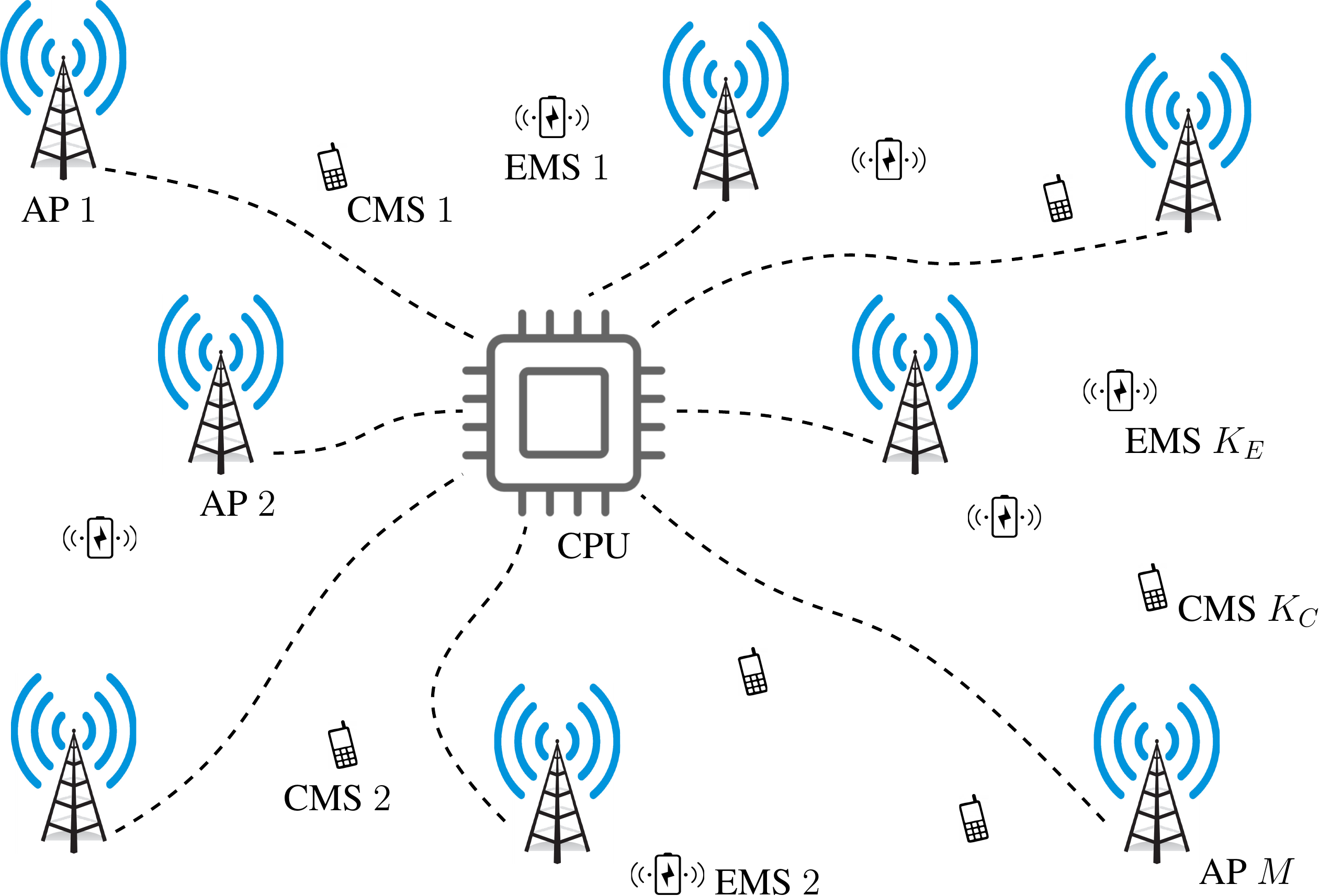}
  \caption{System model for the SWIPT-enhanced cell-free massive MIMO system with $M$ multi-antenna APs connected to a \gls{CPU} via perfect fronthaul links and simultaneously serving $K_C$ CMSs and $K_E$ EMSs.}
  \label{fig:Network}
\end{figure}

\section{System model}
\label{sec:System_model}

\begin{figure}
  \centering
  \includegraphics[width=8.5cm]{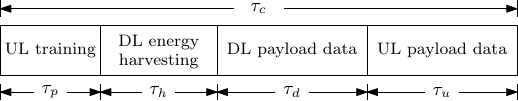}
  \caption{\Gls{TDD} frame of the proposed \gls{SWIPT}-enhanced \gls{CF-mMIMO} network.}\label{fig:TDD_SWIPT}
\end{figure}

As shown in Fig.~\ref{fig:Network}, this work deals with a \gls{SWIPT}-enhanced \gls{CF-mMIMO} network comprising $M$ \glspl{AP}, all connected to a \gls{CPU} via perfect fronthaul links (i.e, error free links with infinite-capacity), and providing service to $K_E$ \glspl{EMS} and $K_C$ \glspl{CMS}. Both \glspl{EMS} and \glspl{CMS} will be indexed by the sets $\mathcal{K}_E=\{1,\ldots,K_E\}$ and $\mathcal{K}_C=\{K_E+1,\ldots,K\}$, respectively, where $K=K_E+K_C$.
Furthermore, the set of $K$ \glspl{MS} (including both \glspl{EMS} and \glspl{CMS}) will be indexed by the set $\mathcal{K}=\mathcal{K}_E \cup \mathcal{K}_C=\{1,\ldots,K\}$.
Each \gls{AP} is equipped with an array of $N$ antennas and the \glspl{CMS} and \glspl{EMS} are single-antenna devices.
A \gls{TDD} protocol is used to organize the transmissions between \glspl{AP} and \glspl{MS} through which each coherence interval is partitioned into four phases: the \gls{UL} training phase, the \gls{DL} energy harvesting phase, the \gls{DL} payload data transmission phase and the \gls{UL} payload data transmission phase (see Fig.~\ref{fig:TDD_SWIPT}).
During the \gls{UL} training, all \glspl{MS} transmit training pilots, thus enabling the \glspl{AP} to estimate the propagation channels to each \gls{MS}. These channel estimates are then used to efficiently transfer power in the \gls{DL} energy harvesting phase, to design the precoding filters used for the \gls{DL} payload data transmission, and to detect the signals sent from the \glspl{MS} in the \gls{UL} payload data transmission phase.
The duration/bandwidth of the combination of training, energy harvesting, \gls{DL} and \gls{UL} phases, denoted as $\tau_p$, $\tau_h$, $\tau_d$ and $\tau_u$, respectively, can not exceed the channel's coherence interval, denoted as $\tau_c$ (i.e., $\tau_p+\tau_h+\tau_d+\tau_u\leq \tau_c$), with all these intervals being expressed in samples (also known as \emph{channel uses}) on a time-frequency plane.

Although the \glspl{EMS} are assumed to be equipped with a battery, it is worth pointing out at this point that we are interested in implementing and analyzing energy consumption strategies that, whenever possible, only consume the average energy they have collected to transmit pilots and data. Thus, battery power will only be intended to provide support to internal device operations. In fact, even though it is inevitable that under some particular circumstances the use of battery power will have to be resorted to for transmission (e.g., initial pilot transmission phase, or blockage period), this paper only considers scenarios characterized with a negligible probability that the average harvested energy is not sufficient to carry out these operations.

\subsection{Channel model}

Similar to \cite{Demir20LSFD}, a slightly modified version of the \gls{3GPP} indoor hotspot (InH) channel model described in \cite{3GPP36814} will be used in this work. In particular, the link between the $m$th \gls{AP} and the $k$th \gls{MS} can be either in \gls{LOS} or in \gls{NLOS}, with an \gls{AP}-to-\gls{MS} distance-dependant probability to be in \gls{LOS} denoted as $p_{\text{LOS}}(d_{mk})$, where $d_{mk}$ is the distance between the \gls{AP} and the \gls{MS}, and which can be calculated using the equations specified in \cite[Table B.1.2.1-2]{3GPP36814}. Furthermore, denoting by $\chi_{mk} \sim \mathcal{N}\left(0,\sigma_\chi^2\right)$ the fading caused by shadowing, the propagation losses (on a dB-scale) experienced on the previous link will be characterized as
\begin{equation}
   L_{mk}=\alpha+10 \beta\log_{10}(d_{mk})+\chi_{mk},
\end{equation}
with the specific values for $\alpha$, $\beta$ and $\sigma_\chi$ depending on whether the specific link is subject to \gls{NLOS} or \gls{LOS} propagation. Ngo \emph{et al.} in \cite[(54)-(55)]{Ngo17} provide the characteristics of the spatial correlation used to model this shadow fading.

Comprising both large-scale propagation losses and small-scale fading, the channel between the $k$th \gls{MS} and the $m$th \gls{AP} over an arbitrary coherence interval will be modeled as
\begin{equation}
   \bs{g}_{mk}=\sqrt{\frac{K_{mk}}{K_{mk}+1}}\overline{\bs{h}}_{mk}+\sqrt{\frac{1}{K_{mk}+1}}\bs{h}_{mk},
\end{equation}
with a \gls{LOS} component
\begin{equation}
   \overline{\bs{h}}_{mk}=\alpha_{mk} \text{ } a^{\gls{MS}}\left(\overline{\theta}_{mk,1}^{\gls{MS}},\overline{\phi}_{mk,1}^{\gls{MS}}\right)\bs{a}^{\gls{AP}}\left(\overline{\theta}_{mk,1}^{\gls{AP}},\overline{\phi}_{mk,1}^{\gls{AP}}\right),
\end{equation}
on top of a scattered multipath \gls{NLOS} component
\begin{equation}
\begin{split}
   \bs{h}_{mk} &= \sum_{c=1}^{C_{mk}}\sum_{p=1}^{P_{mk}} \alpha_{mk,cp} \text{ } a^{\gls{MS}}\left(\theta_{mk,cp}^{\gls{MS}},\phi_{mk,cp}^{\gls{MS}}\right) \\
               &\qquad\qquad\qquad\times \bs{a}^{\gls{AP}}\left(\theta_{mk,cp}^{\gls{AP}},\phi_{mk,cp}^{\gls{AP}}\right).
\end{split}
\end{equation}
The Ricean $K$-factor $K_{mk}$ can be obtained as $10 \log_{10}(K_{mk}) \sim \mathcal{N}\left(\mu_K,\sigma_K^2\right)$ for \gls{LOS} propagation links and as $K_{mk}=0$ for \gls{NLOS} links. The propagation losses-dependant complex channel gain of the \gls{LOS} component is given by $\alpha_{mk}=10^{-L_{mk}/20} e^{j\kappa_{mk}}$, with $\kappa_{mk} \sim \mathcal{U}[0, 2\pi]$. The parameters $C_{mk}$ and $P_{mk}$ represent, respectively, the number of scattering clusters of the \gls{NLOS} component and the path number for each particular cluster. The complex-valued gain of the $p$th path of cluster $c$ is denoted as $\alpha_{mk,cp}$. Furthermore, $a^{\gls{MS}}\left(\theta,\phi\right)$ and $\bs{a}^{\gls{AP}}\left(\theta,\phi\right)$ are used to denote, respectively, the \gls{MS} antenna element response and the \gls{AP} array response vector at the generic azimuth and elevation angles $\theta$ and $\phi$. As suggested by the \gls{3GPP} in \cite{3GPP36814} (see also \cite{3GPP17},\cite{Akdeniz14}), the azimuth angles $\theta_{mk,cp}^{\gls{MS}}$ and $\theta_{mk,cp}^{\gls{AP}}$ can be modelled as wrapped Gaussian random variables around the azimuthal center of the cluster $\overline{\theta}_{mk,c}^{\gls{MS}}$ and $\overline{\theta}_{mk,c}^{\gls{AP}}$ with a standard deviation determined by the angular spreads for the cluster. Moreover, the elevation angles $\phi_{mk,cp}^{\gls{MS}}$ and $\phi_{mk,cp}^{\gls{AP}}$ can be generated as Laplacians around the central elevation angles $\overline{\phi}_{mk,c}^{\gls{MS}}$ and $\overline{\phi}_{mk,c}^{\gls{AP}}$ of the cluster with scale parameters determined by the corresponding angular spreads for the cluster. The cluster central azimuthal angles $\overline{\theta}_{mk,c}^{\gls{MS}}$ and $\overline{\theta}_{mk,c}^{\gls{AP}}$ are both distributed as $\mathcal{U}[-\pi,\pi]$ and the cluster central elevation angles $\overline{\phi}_{mk,c}^{\gls{MS}}$ and $\overline{\phi}_{mk,c}^{\gls{AP}}$ coincide with the corresponding \gls{LOS} elevation angles. The angular spreads of the cluster (in the \gls{rms} sense) follow an exponential distribution with mean $1/\lambda_{\gls{rms}}$ that depends on whether the direction under consideration is the azimuth or the elevation. Finally, the complex gain of the $p$th path of cluster $c$ conforming the \gls{NLOS} component of the channel between the $k$th \gls{MS} and the $m$th \gls{AP} is distributed as $\alpha_{mk,cp} \sim \mathcal{CN}\left(0,\gamma_{mk,c}10^{-L_{mk}/10}\right)$, where, based on \cite[eq. (B.10)]{3GPP36814}, $\gamma_{mk,c}=N \gamma'_{mk,c}/\left(P_{mk}\sum_{j=1}^{C_{mk}} \gamma'_{mk,j}\right)$, is the power fraction with which cluster $c$ contributes to the scattered multipath fading, with $\gamma'_{mk,j}=U_{mk,j}^{r_\tau-1} 10^{Z_{mk,j}/10}$ \cite[eq. (7)]{Akdeniz14}, where $U_{mk,j} \sim \mathcal{U}[0,1]$, the variable $Z_{mk,j} \sim \mathcal{N}(0,\zeta^2)$, and model parameters $r_\tau$ and $\zeta^2$ can be found in \cite[Table B.1.2.2.1-4]{3GPP36814}. Assuming the channel propagation model just described, the \gls{LOS} and scattered multipath components can be characterized by the spatial covariance matrices
\begin{equation}
\begin{split}
   \overline{\bs{R}}_{mk} = &\mathbb{E}\left\{\overline{\bs{h}}_{mk} \overline{\bs{h}}_{mk}^H \right\} = 10^{-L_{mk}/10} \left|a^{\gls{MS}}\left(\overline{\theta}_{mk,1}^{\gls{MS}},\overline{\phi}_{mk,1}^{\gls{MS}}\right)\right|^2 \\ &\times \bs{a}^{\gls{AP}}\left(\overline{\theta}_{mk,1}^{\gls{AP}},\overline{\phi}_{mk,1}^{\gls{AP}}\right)\left(\bs{a}^{\gls{AP}}\left(\overline{\theta}_{mk,1}^{\gls{AP}},\overline{\phi}_{mk,1}^{\gls{AP}}\right)\right)^H,
\end{split}
\end{equation}
and
\begin{equation}
\begin{split}
   &\bs{R}_{mk} = \mathbb{E}\left\{\bs{h}_{mk} \bs{h}_{mk}^H \right\} = 10^{-L_{mk}/10}\sum_{c=1}^{C_{mk}}\gamma_{mk,c} \\
   &\quad\times\sum_{p=1}^{P_{mk}} \mathbb{E}\left\{\left|a^{\gls{MS}}\left(\theta_{mk,cp}^{\gls{MS}},\phi_{mk,cp}^{\gls{MS}}\right)\right|^2\right\} \\
   &\quad\times \mathbb{E}\left\{\bs{a}^{\gls{AP}}\left(\theta_{mk,cp},\phi_{mk,cp}\right)\left(\bs{a}^{\gls{AP}}\left(\theta_{mk,cp},\phi_{mk,cp}\right)\right)^H\right\},
\end{split}
\end{equation}
respectively.

\subsection{Channel estimation (training phase)}
\label{subsec:small-scale-training}
The \gls{UL} training phase length (specified in samples on a time-frequency plane) is denoted as $\tau_p$. During this phase, the $K$ \glspl{MS} are assumed to simultaneously transmit pilot sequences of length $\tau_p$ allowing the \glspl{AP} to conduct the channel estimation. The $\tau_p$ samples received on each of the $N$ antennas of the $m$th \gls{AP} can be organized in the $N \times \tau_p$ signal matrix
\begin{equation}
   {\bs{Q}_p}_m=\sum_{k'=1}^K \sqrt{\tau_p {P_p}_{k'}}\bs{g}_{mk'} \bs{\varphi}_{k'}^T+{\bs{N}_p}_m,
\end{equation}
where ${P_p}_k$ is the power used by \gls{MS} $k$ to transmit each of the training symbols, $\bs{\varphi}_k$ corresponds to the $\tau_p\times 1$ pilot sequence allocated to \gls{MS} $k$, assumed to fulfil the constraint $\|\bs{\varphi}_k\|^2=1$, and ${\bs{N}_p}_m$ has independent, identically distributed (i.i.d.) elements that are zero-mean complex Gaussian random variables with variance $\sigma_u^2$. Despite mutually orthogonal pilot sequences should be ideally chosen, in most realistic deployments it will hold that $K>\tau_p$. This implies that different \glspl{MS} may be sharing the same training sequence and, hence, the so-called pilot contamination takes place \cite{Marzetta10,Marzetta16,Elijah16}. Furthermore, in this paper we do not consider the use of pilot power control and thus, a constant per-pilot symbol power ${P_p}_k=P_p^{\gls{CMS}}$ and ${P_p}_k=P_p^{\gls{EMS}}$ is assumed to be allocated to each \gls{CMS} and \gls{EMS}, respectively.

In most scenarios of practical interest, the spatial channel correlation matrices can be assumed to vary at a much slower pace than the small-scale fading components of the channel (i.e., the phase shifts $\kappa_{mk}$ in the \gls{LOS} paths and the complex-valued gains $\alpha_{mk,cp}$ of the \gls{NLOS} components) and, hence, they can be estimated (and tracked) at the $m$th \gls{AP}, for all $k$ \cite{Ngo18Rice}. Under the assumption of a perfect estimation of the large-scale parameters, and as suggested by \"Ozdogan \emph{et al.} in \cite{Ozdogan19}, the impact the estimation of the small-scale terms might have on the performance of the system can be explored by considering the analytically tractable cases where the phase-shift in the \gls{LOS} components are either perfectly known or fully unknown. In both cases, a Bayesian \gls{LMMSE} estimator will be considered, which for the phase-aware estimator case coincides with the \gls{MMSE} estimator. Following the same notation used by \"Ozdogan \emph{et al.} in \cite{Ozdogan19}, the phase-aware estimator will be denoted as \gls{MMSE} and the phase-unaware estimator will be denoted as \gls{LMMSE}. The performance provided by any other practical channel estimator relying on the knowledge of the spatial channel correlation matrices will lie somewhere in between the performance provided by these benchmarking schemes.

Assuming that the expectations are taken over the scattered channel gains $\alpha_{mk,cp}$ for the \gls{MMSE} case and over the phase shifts $\kappa_{mk}$ and the scattered channel gains $\alpha_{mk,cp}$ for the \gls{LMMSE} case, let us define
\begin{equation}
\begin{split}
   \breve{\bs{g}}_{mk}=\bs{g}_{mk}-\mathbb{E}\left\{\bs{g}_{mk}\right\}=\begin{cases}
                                                                           \sqrt{\frac{1}{K_{mk}+1}}\bs{h}_{mk}, & \text{\gls{MMSE}} \\
                                                                           \bs{g}_{mk},                          & \text{\gls{LMMSE}},
                                                                        \end{cases}
\end{split}
\end{equation}
and
\begin{equation}
\begin{split}
    &{\bs{q}_p}_{mk}=\left({\bs{Q}_p}_m-\mathbb{E}\left\{{\bs{Q}_p}_m\right\}\right)\bs{\varphi}_k^* \\
    &\ =\begin{cases}
           \displaystyle\sum_{k'=1}^K \sqrt{\frac{\tau_p {P_p}_{k'}}{K_{mk'}+1}}\bs{h}_{mk'}\bs{\varphi}_{k'}^T\bs{\varphi}_k^*+{\bs{N}_p}_m\bs{\varphi}_k^*, & \text{\gls{MMSE}} \\
           \displaystyle\sum_{k'=1}^K \sqrt{\tau_p {P_p}_{k'}}\bs{g}_{mk'}\bs{\varphi}_{k'}^T\bs{\varphi}_k^*+{\bs{N}_p}_m\bs{\varphi}_k^*, & \text{\gls{LMMSE}},
        \end{cases}
\end{split}
\end{equation}
which can then be used to compute the estimate for the channel between \gls{MS} $k$ and the $m$th \gls{AP} as \cite{Kay93,Ngo18Rice,Ozdogan19}
\begin{equation}
\begin{split}
   &\hat{\bs{g}}_{mk}=\resizebox{.42\textwidth}{!}{$\mathbb{E}\left\{\bs{g}_{mk}\right\}+\mathbb{E}\left\{\breve{\bs{g}}_{mk}{\bs{q}_p}_{mk}^H\right\}\left(\mathbb{E}\left\{{\bs{q}_p}_{mk}{\bs{q}_p}_{mk}^H\right\}\right)^{-1}{\bs{q}_p}_{mk}$} \\
   &\ =\begin{cases}
         \sqrt{\frac{K_{mk}}{K_{mk}+1}}\overline{\bs{h}}_{mk}+\frac{\sqrt{\tau_p {P_p}_k}}{K_{mk}+1} \bs{R}_{mk}\bs{\Psi}_{mk}^{-1} {\bs{q}_p}_{mk}, & \text{\gls{MMSE}} \\
         \sqrt{\tau_p {P_p}_k} \bs{C}_{mk}\bs{\Theta}_{mk}^{-1} {\bs{q}_p}_{mk}, & \text{\gls{LMMSE}}
      \end{cases}
\end{split}
\end{equation}
where
\begin{equation}
   \bs{\Psi}_{mk}=\sum_{k'=1}^K \frac{\tau_p {P_p}_{k'}}{K_{mk'}+1}\bs{R}_{mk'} \left|\bs{\varphi}_{k'}^H\bs{\varphi}_k\right|^2+\sigma_u^2 \bs{I}_N,
\end{equation}
\begin{equation}
   \bs{C}_{mk}=\frac{K_{mk}}{K_{mk}+1}\overline{\bs{R}}_{mk}+\frac{1}{K_{mk}+1}\bs{R}_{mk},
\end{equation}
and
\begin{equation}
   \bs{\Theta}_{mk}=\sum_{k'=1}^K \tau_p {P_p}_{k'}\bs{C}_{mk'} \left|\bs{\varphi}_{k'}^H\bs{\varphi}_k\right|^2+\sigma_u^2 \bs{I}_N.
\end{equation}
As the channel estimation error $\tilde{\bs{g}}_{mk}=\bs{g}_{mk}-\hat{\bs{g}}_{mk}$ and the channel estimate $\hat{\bs{g}}_{mk}$ are uncorrelated random vectors, the spatial covariance matrix of the channel estimation error can be obtained as
\begin{equation}
\begin{split}
    \bs{A}_{mk}&=\mathbb{E}\left\{\tilde{\bs{g}}_{mk}\tilde{\bs{g}}_{mk}^H\right\} \\
               &=\begin{cases}
                    \frac{\bs{R}_{mk}}{K_{mk}+1} - \frac{\tau_p {P_p}_k \bs{R}_{mk}\bs{\Psi}_{mk}^{-1}\bs{R}_{mk}}{\left(K_{mk}+1\right)^2}, & \text{\gls{MMSE}} \\
                    \bs{C}_{mk} - \tau_p {P_p}_k \bs{C}_{mk} \bs{\Theta}_{mk}^{-1} \bs{C}_{mk}, & \text{\gls{LMMSE}}.
                 \end{cases}
\end{split}
\label{eq:tildeAmk}
\end{equation}

\subsection{Downlink energy harvesting phase}

Let us denote by $\bs{e}_d=\left[{e_d}_1 \ldots {e_d}_{K_E}\right]^T$ the $K_E \times 1$ vector of \textit{energy symbols} transmitted to the $K_E$ \glspl{EMS} during the \gls{DL} energy harvesting phase, where it is assumed that $E\left\{\bs{e}_d\bs{e}_d^H\right\}=\bs{I}_{K_E}$. Using this definition, the \textit{energy signals} vector transmitted from the $m$th \gls{AP} can be written as
\begin{equation}
   \bs{z}_m=\bs{V}_{d\,m} \bs{\Omega}_m^{1/2} \bs{e}_d,
\end{equation}
where $\bs{V}_{d\,m}=\left[\bs{v}_{dm1}\ \ldots\ \bs{v}_{dmK_E}\right]\ \in\ \mathbb{C}^{N \times K_E}$ denotes the \textit{energy precoding matrix} at the $m$th \gls{AP}, and the $K_E \times K_E$ diagonal matrix $\bs{\Omega}_m=\diag\left(\bs{\omega}_m\right)=\diag\left([\omega_{m1}\,\ldots\,\omega_{m K_E}]^T\right)$ contains the energy harvesting power weighing coefficients used at the $m$th \gls{AP} on its main diagonal. The power constraints affecting these power control coefficients are
\begin{equation}
   \mathbb{E}\left\{\left\|\bs{z}_m\right\|^2\right\} = \sum_{k=1}^{K_E} \omega_{mk} \zeta_{mk} \leq P_d,
\label{eq:Power_constraint_EMS}
\end{equation}
for all $m\in \mathcal{M}$, where $P_d$ is used to denote the maximum average transmit power available at any of the \glspl{AP} in the network, and
\begin{equation}
   \zeta_{mk}=\mathbb{E}\left\{\left\|\bs{v}_{dmk}\right\|^2\right\}.
\end{equation}
The signal received by \gls{EMS} $k$ can now be written as
\begin{equation}
\begin{split}
   {r_d}_k&=\sum_{m\in \mathcal{M}} \bs{g}_{mk}^T \bs{z}_m + {n_d}_k=\sum_{m\in \mathcal{M}} \bs{g}_{mk}^T \bs{V}_{d\,m} \bs{\Omega}_m^{1/2} \bs{e}_d + {n_d}_k,
\end{split}
\end{equation}
where ${n_d}_k \sim \mathcal{CN}(0,\sigma_d^2)$. Thus, the vector $\bs{r}_d = \left[{r_d}_1\, \ldots\, {r_d}_{K_E}\right]^T$ containing the signals received by the $K_E$ \glspl{EMS} is given by
\begin{equation}
   \bs{r}_d=\sum_{m\in \mathcal{M}} {\bs{G}_E}_m^T \bs{z}_m + \bs{n}_d = {\bs{G}_E}^T \bs{V}_d \bs{\Omega}^{1/2} \bs{e}_d + \bs{n}_d,
\label{eq:rd}
\end{equation}
where the equivalent \gls{MIMO} channel matrix between the $K_E$ \glspl{EMS} and the $M$ \glspl{AP} is $\bs{G}_E=[{\bs{G}_E}_1^T\,\ldots\,{\bs{G}_E}_M^T]^T$, with ${\bs{G}_E}_m=\left[\bs{g}_{m1}\,\ldots\,\bs{g}_{mK_E}\right]$, and the precoding filter and energy harvesting power weighing matrix implemented at the $M$ \glspl{AP} can be expressed as $\bs{V}_d=[\bs{V}_{d\,1}^T\,\ldots\,\bs{V}_{d\,M}^T]^T$ and $\bs{\Omega}=\blockdiag\left(\bs{\Omega}_1\,\ldots\,\bs{\Omega}_M\right)$, respectively.

Even though this energy transfer approach could be eventually adapted to any linear precoding scheme, as far as we know, only the centralized \gls{ZF} multiuser-\gls{MIMO} precoder provides us with a simple and practical solution to the coupled \gls{UL}/\gls{DL} optimization problem that will be posed in the subsequent subsections. In particular, the use of a distributed \gls{CB} \gls{MIMO} precoder to beamform the energy symbols makes the structure of the resulting optimization problem unsolvable through the use of conventional convex optimization tools\footnote{When using a distributed CB MIMO precoder to beamform the energy symbols, the signal received by EMS $k$ can be expressed as
\begin{equation*}
   {r_d}_k=\sum_{m\in \mathcal{M}} \sum_{k'=1}^{K_E} \sqrt{\omega_{mk'}} \bs{g}_{mk}^T \hat{\bs{g}}_{mk'}^* {e_d}_{k'} + {n_d}_k.
\end{equation*}
The average harvested energy at the $k$th EMS can then be written as
\begin{equation*}
\begin{split}
   {E_{h}}_k(\bs{\omega})=&\eta_h \tau_h \sum_{k'=1}^{K_E} \left(\sum_{m\in\mathcal{M}} \omega_{mk'} \Var\left\{\bs{g}_{mk}^H\hat{\bs{g}}_{mk'}\right\}\right. \\
                          &\qquad\qquad\left. + \left(\sum_{m\in\mathcal{M}} \omega_{mk'}^{1/2} \mathbb{E}\left\{\left|\bs{g}_{mk}^H\hat{\bs{g}}_{mk'}\right|^2\right\}\right)^2\right),
\end{split}
\end{equation*}
which is not a linear expression in terms of the components of $\bs{\omega}_k$. A solution to this problem could be to transmit uncorrelated energy symbols from each of the APs in the system. The amount of harvested energy would suffer, however, a dramatic decrease in the scenarios under investigation, with quantities well below those measured using the centralized ZF scheme.}, thus constituting an interesting thread for further research.

Under the centralized \gls{ZF} approach both the power allocation and precoding processes are performed at the \gls{CPU} and thus, all \glspl{AP} transmit using the same power allocation matrix $\bs{\Omega}_1=\ldots=\bs{\Omega}_M=\diag\left(\left[\omega_1 \ldots \omega_{K_E}\right]^T\right)$ and the precoder is obtained as
\begin{equation}
   \bs{V}_d=\hat{\bs{G}}_E^*\left(\hat{\bs{G}}_E^T \hat{\bs{G}}_E^*\right)^{-1}
\end{equation}
where it has been assumed that $\bs{G}_E=\hat{\bs{G}}_E+\tilde{\bs{G}}_E$. Consequently, the signal received by the $k$th \gls{EMS} can be rewritten as
\begin{equation}
\begin{split}
   {r_d}_k = &\bs{g}_k^T \hat{\bs{G}}_E^*\left(\hat{\bs{G}}_E^T \hat{\bs{G}}_E^*\right)^{-1} \bs{\Omega}^{1/2} \bs{e}_d + {n_d}_k \\
           = &\sqrt{\omega_k} {e_d}_k + \tilde{\bs{g}}_k^T \bs{V}_d \bs{\Omega}^{1/2} \bs{e}_d + {n_d}_k,
\end{split}
\label{eq:rdk}
\end{equation}
where we have defined $\bs{g_k}=\left[\bs{g}_{1k}^T\,\ldots\,\bs{g}_{M k}^T\right]^T$. The average harvested energy at the $k$th \gls{EMS} can be derived from \eqref{eq:rdk} as
\begin{equation}
\begin{split}
   {E_{h}}_k(\bs{\omega})=&\eta_h \tau_h \left(\omega_k+\mathbb{E}\left\{\bs{e}_d^H \bs{\Omega}^{1/2} \bs{V}_d^H \tilde{\bs{g}}_k^* \tilde{\bs{g}}_k^T \bs{V}_d \bs{\Omega}^{1/2} \bs{e}_d\right\}\right) \\
            =&\eta_h \tau_h \left(\omega_k+ \sum_{k'=1}^{K_E} \omega_{k'} \vartheta_{kk'}\right),
\end{split}
\label{eq:Ehk}
\end{equation}
where $\eta_h$ is the rectenna efficiency, and we have used the definition
\begin{equation}
   \vartheta_{kk'}=\left[\diag\left(\mathbb{E}\left\{\bs{V}_d^H \bs{A}_k \bs{V}_d\right\}\right)\right]_{k'},
\end{equation}
with $\bs{A}_k=\blockdiag\left[\bs{A}_{1k}\,\ldots\,\bs{A}_{Mk}\right]$, where the expectation must be estimated through Monte Carlo simulations \cite{Nayebi17}.

\subsection{Uplink payload data transmission}

The $K_E$ \glspl{EMS} use the harvested energy to transmit the corresponding \gls{UL} payload data and, also, to transmit the pilot corresponding to the next \gls{UL} training phase. The $K_C$ \glspl{CMS}, in contrast, have an available maximum average transmit power equal to $P_u^{\gls{CMS}}$ and use \gls{UL} power control coefficients $\psi_k$, for all $k\in\mathcal{K}_C$, with $0 \leq \psi_k \leq 1$. The bank of $N$ \gls{RF} chains implemented at the $m$th \gls{AP} provide the vector of received signals
\begin{equation}
\begin{split}
   {\bs{r}_u}_m&=\sum_{k\in\mathcal{K}_E} \sqrt{P_{u\,k}^{\gls{EMS}}(\bs{\omega})}\, \bs{g}_{mk} {s_u}_k \\
               &\quad+ \sum_{k\in\mathcal{K}_C} \sqrt{P_u^{\gls{CMS}} \psi_k}\, \bs{g}_{mk} {s_u}_k + {\bs{n}_u}_m \\
               &= \bs{G}_m \bs{P}_u^{1/2}(\bs{\omega},\bs{\psi}) \bs{s}_u + {\bs{n}_u}_m,
\end{split}
\label{eq:rum}
\end{equation}
where $\bs{G}_m=\left[\bs{g}_{m1}\,\ldots\,\bs{g}_{mK}\right]$ is the channel matrix between the $K$ \glspl{MS} and the $m$th \gls{AP}, $\bs{P}_u(\bs{\omega},\bs{\psi})=\diag([{P_u}_1(\bs{\omega},\psi_1)\,\ldots\,{P_u}_K(\bs{\omega},\psi_K)]^T)$, with
\begin{equation}
   {P_u}_k(\bs{\omega},\psi_k)=\begin{cases}
                                     P_{u\,k}^{\gls{EMS}}(\bs{\omega}), & k\in\mathcal{K}_E \\
                                     P_u^{\gls{CMS}} \psi_k, & k\in\mathcal{K}_C,
                                  \end{cases}
\label{eq:Puk}
\end{equation}
is a diagonal matrix containing the transmit powers of the $K$ \glspl{MS} on its main diagonal, $\bs{s}_u=[{s_u}_1\,\ldots\,{s_u}_K]^T$ is a vector containing the set of symbols transmitted by the $K$ \glspl{MS}, with $E\left\{\bs{s}_u\bs{s}_u^H\right\}=\bs{I}_K$, and ${\bs{n}_u}_m \sim \mathcal{CN}(\bs{0},\sigma_u^2 \bs{I}_N)$ is the vector of additive noise samples. Note that under the assumption that the energy harvested by \gls{EMS} $k$ is sufficient to perform both the \gls{UL} payload data transmission in this coherence interval and the \gls{UL} pilot transmission in the next coherence interval, the corresponding \gls{UL} payload transmitted power can be expressed as
\begin{equation}
   P_{u\,k}^{\gls{EMS}}(\bs{\omega})=\frac{\alpha_k^{\gls{EMS}} E_{h\,k}(\bs{\omega}) - \tau_p P_p^{\gls{EMS}}}{\tau_u},
\label{eq:PukEMS}
\end{equation}
where $\alpha_k^{\gls{EMS}}$ is the efficiency of the power amplifier at the $k$th \gls{EMS}. That is, a part of the harvested energy is dissipated in the power amplifier, a further portion is reserved to transmit the \gls{UL} pilot in the next coherence interval, and only the remaining energy can be used to transmit the \gls{UL} payload data. In this research work, it has been assumed a negligible probability of $P_{u\,k}^{\gls{EMS}}(\bs{\omega})$ being negative (note that in the unlikely event that this happens, it would not be possible to transmit \gls{UL} payload data or, if necessary, the \gls{EMS} should resort to the use of the battery in order to be able to transmit both pilot and data symbols).

Making use of the fronthaul links, the received signal vectors at the \glspl{AP} are sent to the \gls{CPU} where, as a major difference with respect to previous research work on this topic \cite{Demir20maxmin,Demir20LSFD,Wang20WPT}, they are jointly processed. In fact, the matrix $\bs{W}_u=\left[{\bs{W}_u}_1\,\ldots\,{\bs{W}_u}_M\right]$ is used at the \gls{CPU} to jointly process the vector $\bs{r}_u=\left[{\bs{r}_u^T}_1\,\ldots\,{\bs{r}_u^T}_m\right]^T$ and obtain
\begin{equation}
\begin{split}
   \bs{y}_u=&\bs{W}_u \bs{r}_u=\bs{W}_u \bs{G} \bs{P}_u^{1/2}(\bs{\omega},\bs{\psi}) \bs{s}_u + \bs{\eta}_u,
\end{split}
\end{equation}
where $\bs{G}=\left[\bs{G}_1^T \ldots \bs{G}_M^T\right]^T$, and
\begin{equation}
   \bs{\eta}_u=\bs{W}_u\bs{n}_u=\bs{W}_u \left[{\bs{n}_u^T}_1\,\ldots\,{\bs{n}_u^T}_M\right]^T.
\end{equation}

When using \gls{ZF} \gls{MIMO} detection, the detection matrix used at the \gls{CPU} can be expressed as
\begin{equation}
   \bs{W}_u=\left(\hat{\bs{G}}^H \hat{\bs{G}}\right)^{-1}\hat{\bs{G}}^H.
\end{equation}
Hence, the vector of detected samples can be rewritten as
\begin{equation}
\begin{split}
   \bs{y}_u=\bs{P}_u^{1/2}(\bs{\omega},\bs{\psi}) \bs{s}_u + \bs{W}_u \tilde{\bs{G}} \bs{P}_u^{1/2}(\bs{\omega},\bs{\psi}) \bs{s}_u + \bs{\eta}_u.
\end{split}
\end{equation}
The detected sample corresponding to the transmitted symbol $s_k$ can subsequently be derived as
\begin{equation}
   {y_u}_k =\sqrt{{P_u}_k(\bs{\omega},\psi_k)} {s_u}_k + \left[\bs{W}_u \tilde{\bs{G}} \bs{P}_u^{1/2}(\bs{\omega},\bs{\psi}) \bs{s}_u\right]_k + {\eta_u}_k,
\label{eq:yuk}
\end{equation}
where the first term corresponds to the received useful signal, the interference caused by relying on imperfect \gls{CSI} is represented by the second term, and the thermal noise contribution is represented by the third term.

\section{UL/DL max-min weighted SINR optimization}
\label{sec:UL_DL_max-min}
Analytical procedures resembling the ones in \cite{Hassibi03,Yang13,Interdonato16,Marzetta16,Ngo17,Nayebi17} can be used to derive the \gls{UL} \textit{achievable} \glspl{SINR}. In particular, as the interference and thermal noise terms in \eqref{eq:yuk} are mutually uncorrelated and uncorrelated with the desired signal, they can be treated as \textit{effective noise} allowing the \gls{UL} \textit{achievable} \glspl{SINR} to be expressed as%\footnote{This result relies on the fact that, from a capacity point of view, uncorrelated Gaussian noise samples constitute a worst case scenario. Furthermore, the statistical independence between the complex-valued fast fading random variables characterizing the propagation channels for different \gls{AP}-\gls{MS} pairs is also exploited.}
\begin{equation}
   {\SINR_u}_k(\bs{\omega},\bs{\psi})=\frac{{P_u}_k(\bs{\omega},\psi_k)}{\sum_{k'=1}^K {P_u}_{k'}(\bs{\omega},\psi_{k'}) \delta_{kk'} + \sigma_{\eta_{uk}}^2},
   \label{eq:SINRuk}
\end{equation}
where $\delta_{kk'}=\left[\diag\left(\bs{\Delta_{k'}}\right)\right]_k$, with
\begin{equation}
\begin{split}
   \bs{\Delta_{k'}}&=\mathbb{E}\left\{\bs{W}_u \mathbb{E}\left\{\tilde{\bs{g}}_{k'} \tilde{\bs{g}}_{k'}^H\right\}\bs{W}_u^H\right\}=\mathbb{E}\left\{\bs{W}_u \bs{A}_{k'} \bs{W}_u^H\right\},
\end{split}
\end{equation}
and
\begin{equation}
   \sigma_{\eta_{uk}}^2=\sigma_u^2 \left[\diag\left(\mathbb{E}\left\{\bs{W}_u \bs{W}_u^H\right\}\right)\right]_k.
\end{equation}

Using \eqref{eq:SINRuk}, a coupled UL/DL optimization problem can be formulated aiming at finding the power control coefficients $\omega_k$, for all $k \in \mathcal{K}_E$, used in the \gls{DL} energy harvesting phase, and $\psi_k$, for all $k \in \mathcal{K}_C$, used in the \gls{UL} payload data transmission phase, that maximize the minimum of the weighted achievable \gls{UL} \glspl{SINR} of all \glspl{MS} while satisfying the corresponding \gls{DL} and \gls{UL} transmit power constraints at both the \glspl{AP} and \glspl{MS}. This optimization problem can be formulated as
\begin{equation}
\begin{split}
   &\max_{\{\bs{\omega},\bs{\psi}\}}\ \min_{k\in\{1,\ldots,K\}} \frac{\xi_k {P_u}_k(\bs{\omega},\psi_k)}{\sum_{k'=1}^K {P_u}_{k'}(\bs{\omega},\psi_{k'}) \delta_{kk'} + \sigma_{\eta_{uk}}^2} \\
   &\textrm{subject to } \sum_{k=1}^{K_E} \omega_{mk} \zeta_{mk} \leq P_d,\ \forall m\in\mathcal{M} \\
   &\phantom{\textrm{subject to }} \omega_{mk} \geq 0,\ \forall m\in\mathcal{M}\text{ and }k\in\mathcal{K}_E, \\
   &\phantom{\textrm{subject to }} 0 \leq \psi_k \leq 1,\ \forall k\in\mathcal{K}_C,
\end{split}
\label{eq:opt_problem_UL}
\end{equation}
where $\xi_k$ is the weighting coefficient applied to the \gls{SINR} experienced by \gls{MS} $k$. The use of weighting coefficients $\xi_k$, for $k\in\{1,\ldots,K\}$, allow for the implementation of a basic radio resource management mechanism to control the amount of resources allocated to different \glspl{MS} \cite{Femenias12}. In particular, the higher (with respect to the others) the weighting coefficient allocated to a given \gls{MS}, the lower the relative SINR experienced by that particular user. As the \glspl{EMS} typically transmit a much lower amount of information than the \glspl{CMS}, the allocation of very dissimilar weighting coefficients to both sets of users can be interpreted as a straightforward strategy to control their disparate performance requirements.

Problem \eqref{eq:opt_problem_UL} is a quasi-linear optimization program that can be expressed in an equivalent form as
\begin{equation}
\begin{split}
   &\max_{\{\bs{\omega},\bs{\psi},x\}} x \\
   &\textrm{s. t. } x \left(\sum_{k'=1}^K {P_u}_{k'}(\bs{\omega},\psi_{k'}) \delta_{kk'} + \sigma_{\eta_{uk}}^2\right) \\
   &\phantom{\textrm{s. t. }} \qquad\qquad\qquad\leq \xi_k {P_u}_k(\bs{\omega},\psi_k), \ \forall\,k \in \mathcal{K}, \\
   &\phantom{\textrm{s. t. }} \sum_{k=1}^{K_E} \omega_{mk} \zeta_{mk} \leq P_d,\ \forall m\in\mathcal{M} \\
   &\phantom{\textrm{s. t. }} \omega_{mk} \geq 0,\ \forall m\in\mathcal{M}\text{ and }k\in\mathcal{K}_E, \\
   &\phantom{\textrm{s. t. }} 0 \leq \psi_k \leq 1,\ \forall k\in\mathcal{K}_C.
\end{split}
\label{eq:opt_problem_UL_equiv_form}
\end{equation}
This optimization problem can be solved in a very efficient manner using an iterative bisection search technique (see \cite{Ngo17,Nayebi17} for details regarding the feasibility, optimality and complexity when solving this problem). Indeed, a linear feasibility problem is solved at each iteration of the bisection search. Note that the power control coefficients are obtained on the basis of large-scale channel state information and, consequently, they typically vary at the pace of tens of the coherence interval of the channel. Under these conditions, \glspl{AP} can inform the \glspl{CMS} about the optimal power coefficients $\psi_k$ with a negligible consumption of both bandwidth and energy by using the conventional signaling protocols among these entities.

\section{Performance metrics}
\label{sec:Performance_metrics}

Our ultimate objective is to obtain mathematically tractable expressions of the main \gls{UL} network performance metrics in the \gls{SWIPT}-enhanced \gls{CF-mMIMO} context.
Hence, this section will focus on average spectral/energy efficiencies and power consumption expressions.

\subsection{UL spectral efficiency}

The previous max-min SINR optimization analysis can be used to derive the \gls{UL} spectral efficiency (commonly referred to as achievable rate).  The \gls{UL} spectral efficiency of \gls{MS} $k$ (in bits per second per Hz) can be obtained as
\begin{equation}
   {S_e}_{uk}(\bs{\omega},\bs{\psi})=\frac{\tau_u}{\tau_c} \log_2\left(1+{\SINR_u}_k(\bs{\omega},\bs{\psi})\right),
   \label{eq:UL_SEperMS}
\end{equation}
and the average \gls{UL} sum spectral efficiency is given by
\begin{equation}
   {S_e}_u(\bs{\omega},\bs{\psi})=\sum_{k=1}^K {S_e}_{uk}(\bs{\omega},\bs{\psi}).
   \label{eq:UL_SE}
\end{equation}

\subsection{UL power consumption}\label{subsec:ULPowCons}

The UL power consumption model corresponding to the payload transmission stage encompasses the power drained at the different components of the network (i.e., \glspl{AP}, \glspl{MS} and fronthaul links), and it can be approximated by a linear model as follows (see \cite{Auer11,Tombaz11,Desset12,Bjornson16c,Dai16,Nguyen17,Ngo18,femenias2020access,GarciaMorales20} and references therein). Typically, the fronthaul connection linking the $m$th \gls{AP} to the \gls{CPU} consumes a power that exhibits a proportional dependence on the transported amount of data traffic. During the \gls{UL} payload transmission phase, the $m$th fronthaul link has to transport the information contained in the received vectors ${\bs{r}_u}_m$ on the corresponding $\tau_u$ time/frequency samples. From standard random coding arguments \cite{Cover06}, vector $\bs{s}_u$ can be safely assumed to be distributed as $\bs{s}_u\sim\mathcal{CN}(0,\bs{I}_K)$ and thus, when conditioned on $\bs{G}_m$, the received vector is distributed as ${\bs{r}_u}_m|\bs{G}_m\sim\mathcal{CN}\left(\bs{0},\bs{G}_m\bs{P}_u(\bs{\omega},\bs{\psi})\bs{G}_m^H+\sigma_u^2\bs{I}_N\right)$. Furthermore, as the differential entropy of a vector $\bs{x}\sim\mathcal{CN}(\bs{\mu},\bs{\Theta})$ is given by $\mathcal{H}(\bs{x})=\log\det(\pi e \bs{\Theta})$ \cite{Cover06}, the amount of information contained in ${\bs{r}_u}_m$ can be obtained as (in bps/Hz)
\begin{equation}
\begin{split}
   &{C_u}_m(\bs{\omega},\bs{\psi}) = \mathbb{E}\left\{I\left({\bs{r}_u}_m;\bs{G}_m \bs{P}_u^{1/2}(\bs{\omega},\bs{\psi})\bs{s}_u\right)\right\} \\
   &\quad= \mathbb{E}\left\{\mathcal{H}\left({\bs{r}_u}_m\right)\right\}-\mathbb{E}\left\{\mathcal{H}\left({\bs{r}_u}_m\bigr|\bs{G}_m \bs{P}_u^{1/2}(\bs{\omega},\bs{\psi})\bs{s}_u\right)\right\} \\
   &\quad= \mathbb{E}\left\{\log_2\det\left(\frac{\bs{G}_m\bs{P}_u(\bs{\omega},\bs{\psi})\bs{G}_m^H}{\sigma_u^2}+\bs{I}_N\right)\right\},
\end{split}
\end{equation}
where $I(\bs{x};\bs{y})$ is used to denote the mutual information between vectors $\bs{x}$ and $\bs{y}$, and $\mathcal{H}(\bs{x}|\bs{y})$ is the differential entropy of $\bs{x}$ conditioned on $\bs{y}$. Thus, the \gls{UL} power consumption can be approximated as \cite{Nguyen17,Ngo18}
\begin{equation}
   P_m^{\text{FH}}=B \frac{\tau_u}{\tau_c} \varrho_m^{\text{FH}} {C_u}_m(\bs{\omega},\bs{\psi}) + P_m^{\text{FH,fix}},
\end{equation}
where the bandwidth is denoted by $B$, $\varrho_m^{\text{FH}}$ is the traffic-dependent power consumption coefficient (in Watt per bit/s) and $P_m^{\text{FH,fix}}$ models the consumed power part known to be traffic-independent.

The UL power consumed by the $m$th \gls{AP} during the \gls{UL} payload transmission phase will depend, again, on the amount of information received on this particular set of time/frequency samples. Moreover, it will also depend on other system parameters including the baseband processing power, the small-signal \gls{RF} transceiver power, or the DC-DC power supply, feeder, main supply and cooling system losses \cite{Auer11,Tombaz11,Desset12,Bjornson16c,Dai16}. Hence, a fairly accurate model for the total UL power consumed at the $m$th \gls{AP} is
\begin{equation}
   P_m^{\text{AP}}=B \frac{\tau_u}{\tau_c} \varrho_m^{\text{AP}} {C_u}_m(\bs{\omega},\bs{\psi}) + P_{m\,u}^{\text{AP,fix}} + N P_{m\,u}^{\text{AP,chain}},
\end{equation}
where $\varrho_m^{\text{AP}}$ is the traffic-dependent power consumption coefficient (in Watt per bit/s), $P_{m\,u}^{\text{AP,fix}}$ denotes the amount of consumed \gls{UL} power independently of the traffic load and, finally, the traffic-independent \gls{UL} power drained by the circuitry of every \gls{RF} chain at the $m$th \gls{AP} is denoted by $P_{m\,u}^{\text{AP,chain}}$.

The \gls{UL} power consumed by the $k$th \gls{CMS} can also be modelled as
\begin{equation}
   P_k^{\gls{CMS}}=\frac{P_u^{\gls{CMS}} \psi_k}{\alpha_k^{\text{CMS}}} + P_{k\,u}^{\text{CMS,fix}},
\end{equation}
where the efficiency of the power amplifier is denoted as $\alpha_k^{\gls{CMS}}$, and $P_{k\,u}^{\text{CMS,fix}}$ represents the data-independent constant amount of power consumed by the $k$th \gls{CMS}. The \gls{UL} power consumed by the $k$th \glspl{EMS} to perform both the UL payload data transmission in this coherence interval and the UL pilot transmission in the next coherence is assumed to come from harvested energy. However, the \glspl{EMS} need to maintain a fixed power consumption to keep on their internal circuits. Thus, the \gls{UL} power consumed by the $k$th \gls{EMS} can be approximated as
\begin{equation}
   P_k^{\gls{EMS}}=P_{k\,u}^{\gls{EMS},\text{fix}},
\end{equation}
where, again, $P_{k\,u}^{\gls{EMS},\text{fix}}$ models the data-independent constant amount of consumed power.

Grouping the previous results, the overall power consumed by the whole set of components of the \gls{CF-mMIMO} network corresponding to the \gls{UL} payload transmission stage can be expressed as
\begin{equation}
\begin{split}
   {P_T}_u(\bs{\omega},\bs{\psi})=&P_{T u}^{\text{fix}}+\sum_{\forall k \in \mathcal{K}_C} \frac{\tau_u P_u^{\text{CMS}} \psi_k}{\tau_c \alpha_k^{\text{CMS}}} \\
                                  &+ B \frac{\tau_u}{\tau_c} \sum_{m=1}^M \left(\varrho_m^{\text{AP}} + \varrho_m^{\text{FH}}\right) {C_u}_m(\bs{\omega},\bs{\psi}),
\end{split}
\end{equation}
with
\begin{equation}
\begin{split}
    P_{T u}^{\text{fix}}=\frac{\tau_u}{\tau_c}&\left[\sum_{\forall k \in \mathcal{K}_C} P_{k\,u}^{\text{CMS,fix}} + \sum_{\forall k \in \mathcal{K}_E} P_{k\,u}^{\text{EMS,fix}} \right. \\
                                              &\left.\ + \sum_{m=1}^M \left(P_m^{\text{FH,fix}} + P_{m\,u}^{\text{AP,fix}} + N P_{m\,u}^{\text{AP,chain}}\right)\right].
\end{split}
\end{equation}

\subsection{UL energy efficiency}

Being defined as the ratio between the achievable rate and the transmitted power consumption, the \gls{UL} average energy efficiency can be expressed as
\begin{equation}
   {E_e}_u(\bs{\omega},\bs{\psi})=\frac{B {S_e}_u(\bs{\omega},\bs{\psi})}{{P_T}_u(\bs{\omega},\bs{\psi})}.
\end{equation}

\section{Numerical results}
\label{sec:numerical_results}

\begin{table}[t!]
\renewcommand{\arraystretch}{1.1}
\caption{Summary of default simulation parameters}
\label{tab:parameters}
\centering
\begin{tabular}{l|c}
\hline

\bfseries Parameters & \bfseries Value\\

\hline

\footnotesize {Carrier frequency: $f_0$} & \footnotesize {3.4 GHz}\\
\footnotesize {Bandwidth: $B$} & \footnotesize {20 MHz}\\
\footnotesize {Side of the square coverage area: $D$} & \footnotesize {100 m}\\

\footnotesize {AP/MS antenna height: $h_{\text{AP}}/h_{\text{MS}}$} & \footnotesize {6/1.65 m}\\
\footnotesize {AP/MS noise figure:} & \footnotesize {7/9 dB} \\

\footnotesize {AP maximum transmit power: $P_d$} & \footnotesize {200 mW}\\
\footnotesize {MS maximum transmit power: $P_u$} & \footnotesize {0.1 mW}\\
\footnotesize {Pilot transmit power: $P_p$} & \footnotesize {-40 dBm}\\
\footnotesize {Pilot transmit power coefficient: $\alpha_p$} & \footnotesize {1}\\

\footnotesize {Antenna configuration at each AP: $N_x \times N_y$} & \footnotesize {2 $\times$ 2}\\
\footnotesize {Minimum separation between antenna elements} & \footnotesize {$\lambda/2$}\\

\footnotesize {Coherence interval length: $\tau_c$} & \footnotesize {200 samples}\\
\footnotesize {Training phase length: $\tau_p$} & \footnotesize {20 samples}\\
\footnotesize {Energy harvesting phase length: $\tau_h$} & \footnotesize {30 samples}\\
\footnotesize {UL payload transmission phase length: $\tau_u$} & \footnotesize {$(\tau_c-\tau_p-\tau_h)/2$}\\

\footnotesize {Loss parameters Case LOS: $\alpha$, $\beta$, $\sigma_\chi$} & \footnotesize{43.43, 1.69, 3 dB} \\
\footnotesize {Loss parameters Case NLOS: $\alpha$, $\beta$, $\sigma_\chi$} & \footnotesize{22.13, 4.33, 4 dB} \\

\footnotesize {Decorrelation distance of the shadowing: $d_{\text{dcorr}}$} & \footnotesize {9 m}\\
\footnotesize {Correlation among the shadowing at the \glspl{AP}:} & \footnotesize {0.5}\\

\footnotesize {Distribution of the Ricean $K$-factor: $\mu_K$/$\sigma_K$} & \footnotesize {7/4 dB}\\

\footnotesize {Number of clusters LOS/NLOS: $C_{mk}$} & \footnotesize {15/19}\\
\footnotesize {Number of paths per cluster LOS/NLOS: $P_{mk}$} & \footnotesize {20/20}\\

\footnotesize {Azimuth angular spread (AP) LOS/NLOS: $\lambda_{\text{rms}}^{-1}$} & \footnotesize {5$^o$/5$^o$}\\
\footnotesize {Azimuth angular spread (MS) LOS/NLOS: $\lambda_{\text{rms}}^{-1}$} & \footnotesize {8$^o$/11$^o$}\\
\footnotesize {Elevation angular spread: $\lambda_{\text{rms}}^{-1}$} & \footnotesize {7$^o$}\\

\footnotesize {Cluster power fraction parameters: $r_\tau/\zeta$} & \footnotesize {3/4}\\

\footnotesize {Rectenna efficiency: $\eta_h$} & \footnotesize {0.4}\\
\footnotesize {Power amplifier efficiency at the \glspl{CMS}: $\alpha_k^{\text{CMS}}$} & \footnotesize {0.3}\\
\footnotesize {Power amplifier efficiency at the \glspl{EMS}: $\alpha_k^{\text{EMS}}$} & \footnotesize {0.92}\\
\footnotesize {Traffic-dependent coefficients: $\varrho_m^{\text{AP}}/\varrho_m^{\text{FH}}$} & \footnotesize {0.25/0.25 $\frac{\text{W}}{\text{Gbps}}$}\\
\footnotesize {\gls{AP} fixed power:  $P_{m\,u}^{\text{AP,fix}}$} & \footnotesize {6 W}\\
\footnotesize {\gls{AP} fixed power per RF chain: $P_{m\,u}^{\text{AP,chain}}$} & \footnotesize {0.15 W}\\
\footnotesize {\gls{CMS} fixed power: $P_{k\,u}^{\text{CMS,fix}}$} & \footnotesize {0.75 W}\\
\footnotesize {\gls{EMS} fixed power: $P_{k\,u}^{\text{EMS,fix}}$} & \footnotesize {2.7 $\mu$W}\\
\footnotesize {FH fixed power: $P_m^{\text{FH,fix}}$} & \footnotesize {5 W}\\
\hline
\end{tabular}
\end{table}

An extensive collection of numerical results is presented in this section that, on the one hand, provides an exhaustive performance evaluation of the proposed \gls{SWIPT}-enhanced \gls{CF-mMIMO} system and, on the other hand, reveals the existing trade-offs among the design parameters used to configure the network. The system performance is evaluated in terms of channel estimation \gls{NMSE}, spectral efficiency per \gls{MS}, average energy harvesting capabilities per \gls{EMS}, and average energy/spectral efficiencies and power consumption corresponding to the uplink payload transmission stage. To this end, a cell-free scenario is studied where the $M$ \glspl{AP} and $K$ \glspl{MS} are uniformly distributed at random within a square coverage area of side $D$ with wrap-around \cite{Ngo17}. Without loss of generality, and following \cite{Femenias19}, a balanced random pilot allocation scheme has been applied whereby \glspl{MS} are assigned training sequences that are cyclically and sequentially obtained from a set of ordered orthogonal pilots. The per-pilot symbol transmit power has been set to $P_p^{\gls{CMS}}=P_p$ for the \glspl{CMS} and $P_p^{\gls{EMS}} = \alpha_p P_p$ for the \glspl{EMS}, where $\alpha_p$ is a positive real quantity. Furthermore, the \gls{SINR} weighting coefficients $\xi_k$ have been set to $\xi_k=$~1 for all $k \in \mathcal{K}_C$ and $\xi_k=\xi_{k_E}$ for all $k \in \mathcal{K}_E$. Unless otherwise stated, the default simulation parameters used to carry out the performance evaluation will be those summarized in Table \ref{tab:parameters}. Inspired by \cite{Demir20LSFD}, the values of these parameters have been extracted from the 3GPP indoor hotspot (InH) scenario \cite{3GPP36814} and from some of the most relevant research works on \gls{CF-mMIMO} (see, for instance, \cite{Auer11,Bjornson16c,Ngo17,Nguyen17,3GPP17,Bjornson19} and references therein).

\subsection{Impact of the training phase length}\label{subsec:impact_taup}

\begin{figure*}[t!]
    \centering

    \begin{subfigure}[t]{0.328\textwidth}
        \centering
        \includegraphics[height=7.2cm]{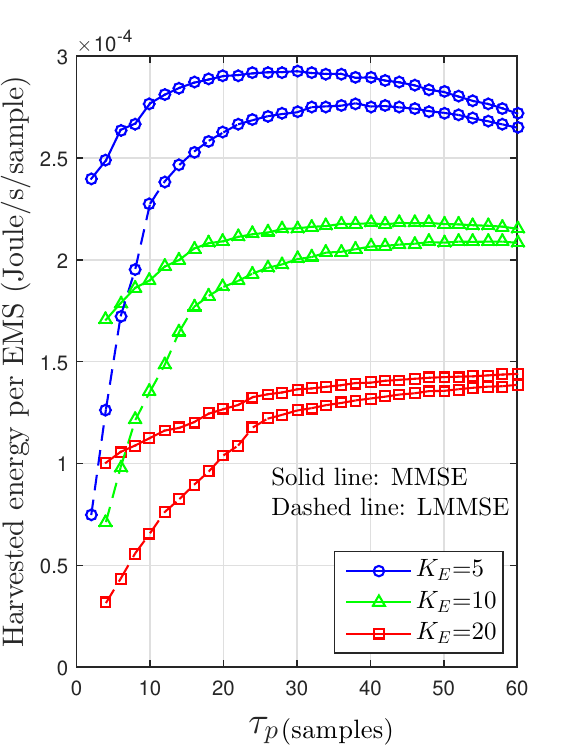}
        \caption{ }
        \label{fig:EhpMS_vs_Tp}
    \end{subfigure}
    \begin{subfigure}[t]{0.328\textwidth}
        \centering
        \includegraphics[height=7.2cm]{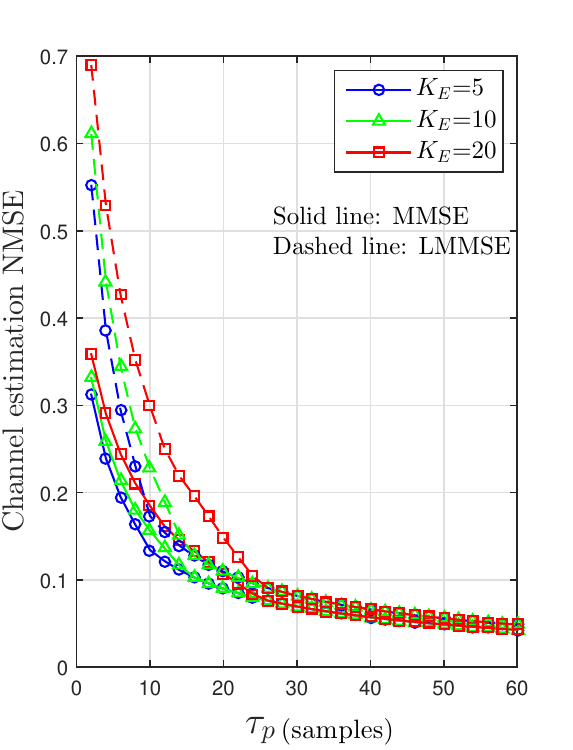}
        \caption{ }
        \label{fig:MMSE_vs_Tp}
    \end{subfigure}
    \begin{subfigure}[t]{0.328\textwidth}
        \centering
        \includegraphics[height=7.2cm]{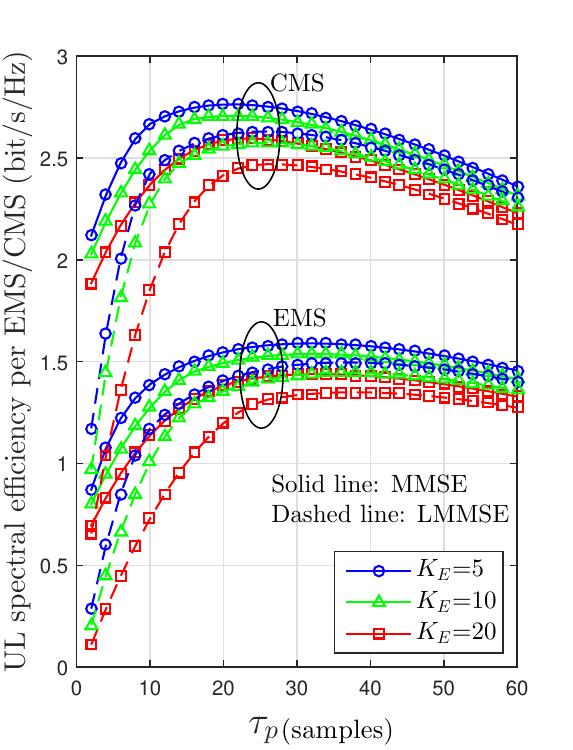}
        \caption{ }
        \label{fig:SepMS_vs_Tp}
    \end{subfigure}

    \caption{Average harvested energy per \gls{EMS}, average channel estimation NMSE, and UL spectral efficiency per MS as a function of the training phase length and using the number of \glspl{EMS} in the network as parameter ($M=$~50 \glspl{AP}, $K_C=$~5 \glspl{CMS}, and $\xi_{k_E}=$~10).}\label{fig:FvsTp}
\end{figure*}

Figure~\ref{fig:FvsTp} shows the average energy harvested by the \glspl{EMS}, the average \gls{NMSE} associated to the channel estimation process, and the \gls{UL} spectral efficiency per \gls{MS} as a function of the training phase length $\tau_p$ and using the number of \glspl{EMS} in the network as parameter. Results have been obtained assuming a \gls{CF-mMIMO} network with $M=$~50 \glspl{AP} serving $K_C=$~5 \glspl{CMS} and using \gls{SINR} weighting coefficients $\xi_{k_C}=$~1 and $\xi_{k_E}=$~10, implying that \glspl{CMS} strive at an \gls{SINR} ten times stronger than \glspl{EMS}. For a given number of \glspl{MS} in the network $K=K_C+K_E$, if $\tau_p < K$ there is a shortage of orthogonal pilot sequences and thus, the system has to cope with very high levels of pilot contamination leading to a large channel estimation \gls{NMSE} and low figures of both the average energy harvested per \gls{EMS} and the spectral efficiency per \gls{MS}. Remarkably, even though the performance of the system based on the \gls{MMSE} channel estimator is considerably affected by the aforementioned shortage of orthogonal pilot sequences, its effects are even more exacerbated when relying on the \gls{LMMSE} estimator. As the length of the training phase increases, up to $\tau_p=K$, there are more available orthogonal pilot sequences in the network allowing for a higher quality channel estimation process that reverts in a performance increase in terms of both the energy harvesting capabilities and spectral efficiency. As shown in Fig.~\ref{fig:MMSE_vs_Tp}, increasing the training phase length beyond $\tau_p=K$ still provides an important improvement in channel estimation quality but, as it can be observed in Fig.~\ref{fig:EhpMS_vs_Tp}, this does not directly result in an improvement of the energy harvesting potential of the \glspl{EMS}. This can be explained based on two main reasons that are somehow connected due to the implementation of the coupled \gls{UL}/\gls{DL} optimization problem in \eqref{eq:opt_problem_UL_equiv_form}. The first one is that the larger the length of the training phase the larger is the portion of harvested energy that the \glspl{EMS} invest in transmitting the \gls{UL} pilot sequence and the smaller is the one they can capitalize on to transmit payload data. The second reason is that, as $\tau_p$ increases, the length of the \gls{UL} payload data transmission phase reduces and this results, as shown in Fig. \ref{fig:SepMS_vs_Tp}, in a decrease in the spectral efficiency provided to the both \glspl{EMS} and \glspl{CMS}. In fact, as the spectral efficiencies provided to \glspl{CMS} and \glspl{EMS} are tightly connected via the coupled \gls{UL}/\gls{DL} max-min optimization problem, this affects the values of the power control coefficients $\bs{\omega}$ that govern the energy that the \glspl{EMS} need to harvest in order to fulfill the optimization constraints (see equation \eqref{eq:Ehk}). Consequently, it can be observed in Fig.~\ref{fig:FvsTp} that the average harvested energy per \gls{EMS} increases with the training phase length $\tau_p$ up to an optimal value that depends on the value of $K_E$ and, moreover, it can be very different depending on whether the objective function is the aggregate spectral efficiency or the spectral efficiency experienced by a particular group of \glspl{MS} (i.e., \glspl{EMS} or \glspl{CMS}). Remarkably, the higher the value of $\tau_p$ the lower is the difference in channel estimation \gls{NMSE} between the systems based on \gls{MMSE} and \gls{LMMSE} channel estimators and, consequently, the lower the relative performance gains provided by the use of the \gls{MMSE} channel estimator.

Another result worth mentioning is that, for a fixed value of $\tau_p$, and irrespective of the channel estimator under evaluation, increasing the number of \glspl{EMS} in the network decreases both the average harvested energy per \gls{EMS} and the spectral efficiency per \gls{MS}. This is mainly due to that the higher the number of \glspl{EMS} in the network, the harder is to design a beamformer able to direct \textit{orthogonal} beams to each of the energy harvesting devices and, moreover, the larger is the set of devices to which the power available at the \glspl{AP} must be allocated during the energy harvesting phase. As expected, the value of $\tau_p$ maximizing the \gls{UL} spectral efficiency per \gls{MS} slightly increases with the number of \glspl{EMS} in the network.

\subsection{Impact of the energy harvesting phase length}\label{subsec:impact_tauh}

\begin{figure*}[t!]
    \centering
    \begin{subfigure}[t]{0.48\textwidth}
        \centering
        \includegraphics[width=\textwidth]{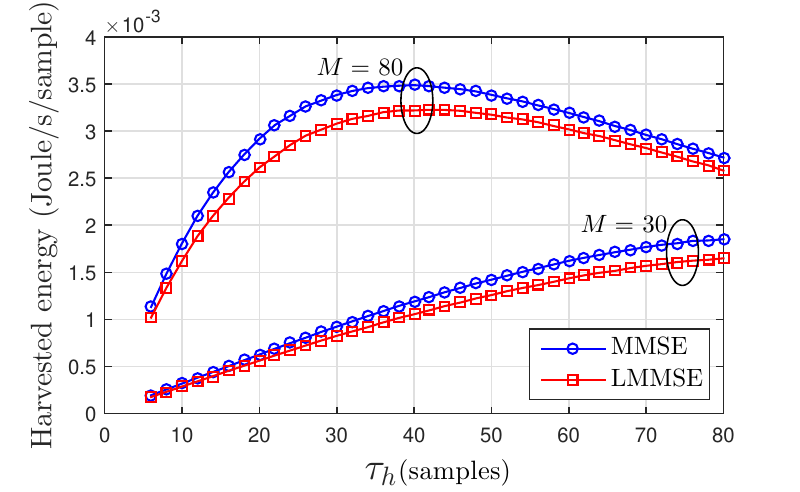}
        \caption{ }
        \label{fig:Eh_vs_Th}
    \end{subfigure}
    \begin{subfigure}[t]{0.48\textwidth}
        \centering
        \includegraphics[width=\textwidth]{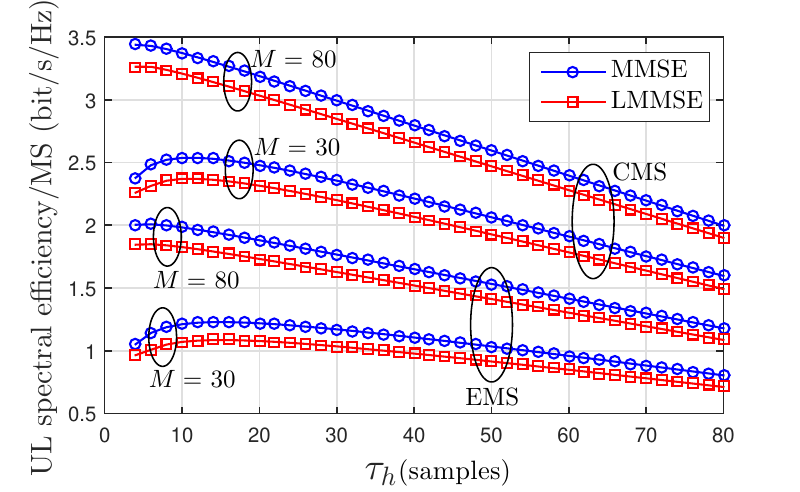}
        \caption{ }
        \label{fig:Se_vs_Th_Ms}
    \end{subfigure}

    \caption{Aggregate harvested energy at the \glspl{EMS} and \gls{UL} spectral efficiency for both \glspl{EMS} and \glspl{CMS} as a function of the energy harvesting phase length and with the number of \glspl{AP} in the network as parameter ($K_C=$~5 \glspl{CMS}, $K_E=$~10 \glspl{EMS}, and $\xi_{k_E}=$~10.)}\label{fig:FvsTh}
\end{figure*}

The impact of the energy harvesting phase length $\tau_h$ on the average harvested energy and the spectral efficiency of both the \glspl{EMS} and \glspl{CMS} is evaluated in Fig.~\ref{fig:FvsTh}. Results have been obtained assuming \gls{CF-mMIMO} networks with different densities of \glspl{AP} per area unit, and serving  $K_C=$~5 \glspl{CMS} and $K_E=$~10 \glspl{EMS} with $\xi_{k_E}=$~10. Note that, as we assume a default training phase length of $\tau_p=20$ samples, all the \glspl{MS} in the network are allocated orthogonal pilot sequences. As shown in Fig.~\ref{fig:Eh_vs_Th}, for low values of the energy harvesting phase length, the aggregate energy harvested by the \glspl{EMS} is linearly proportional to this parameter, especially when working on \gls{CF-mMIMO} networks with a low spatial density of \glspl{AP}. For very large values of $\tau_h$, however, and particularly noticeable for a high spatial density of \glspl{AP}, the rate at which the aggregate average harvested energy increases with $\tau_h$ is not linear anymore and it even decreases for large values of $\tau_h$. As shown in Fig. \ref{fig:Se_vs_Th_Ms}, the increment of $\tau_h$ and the particular behaviour of the available harvested energy at the \glspl{EMS} as a function of this parameter, have an interesting effect on the spectral efficiency of both the \glspl{EMS} and \glspl{CMS}. In particular, irrespective of whether we are dealing with the spectral efficiency provided to \glspl{EMS} or that provided to \glspl{CMS}, there is an \textit{optimal} value of the energy harvesting phase length $\tau_h$ that depends on the spatial density of \glspl{AP} in the network. Notably, this \emph{optimal} value of $\tau_h$ is only slightly (almost negligibly) higher for the \gls{LMMSE} channel estimator than for the \gls{MMSE} one. For very small values of $\tau_h$, the amount of harvested energy available at the \glspl{EMS} is unavoidably low and this reverts in a very low spectral efficiency for the \glspl{EMS}. As the spectral efficiencies experienced by the \glspl{CMS} are tightly connected to those provided to the \glspl{EMS} due to the coupled \gls{UL}/\gls{DL} optimization problem, these \glspl{MS} are also \textit{forced} to experience a low spectral efficiency. As the value of $\tau_h$ increases, so does the average harvested energy at the \glspl{EMS}, thus allowing for an increase of the spectral efficiencies experienced by both \glspl{EMS} and \glspl{CMS}. However, as increasing $\tau_h$ translates into a decrease of $\tau_u$, then there exists a trade-off between the potential increase of spectral efficiency provided by the availability of a high level of harvested energy at the \glspl{EMS} and the dramatic decrease of spectral efficiency induced by the reduction of the \gls{UL} payload transmission phase length.

The higher the number of \glspl{AP} deployed in the coverage area, the higher is the number of available energy sources and the shorter is the average distance between a given \gls{EMS} and the surrounding \glspl{AP} and, consequently, the larger is the amount of harvested energy. The quality of the channel estimates and that of the beamformers also increase as more serving \glspl{AP} are added to the network. In fact, the higher the number of serving \glspl{AP}, the lower the difficulty in separating the corresponding propagation channels and the higher the aggregate quota of power that can be allocated to each of the beams. For a fixed value of $\tau_h$, increasing the amount of harvested energy at the \glspl{EMS} translates into increased figures of the spectral efficiency of both the \glspl{EMS} and \glspl{CMS} but, remarkably, given the logarithmic dependence of spectral efficiency on the achievable \gls{SINR}, the relative increase is noticeably more relevant to \glspl{EMS} than to \glspl{CMS}. Moreover, it is worth pointing out that, due to the aforementioned compromise between an increased amount of harvested energy and a reduced length of the \gls{UL} payload data transmission phase, the value of $\tau_h$ maximizing the spectral efficiency provided to \glspl{EMS} decreases as the number of \glspl{AP} in the network increases.

\subsection{Impact of the pilot transmit power}

\begin{figure*}[t!]
    \centering

        \begin{subfigure}[b]{0.48\textwidth}
        \includegraphics[width=\textwidth]{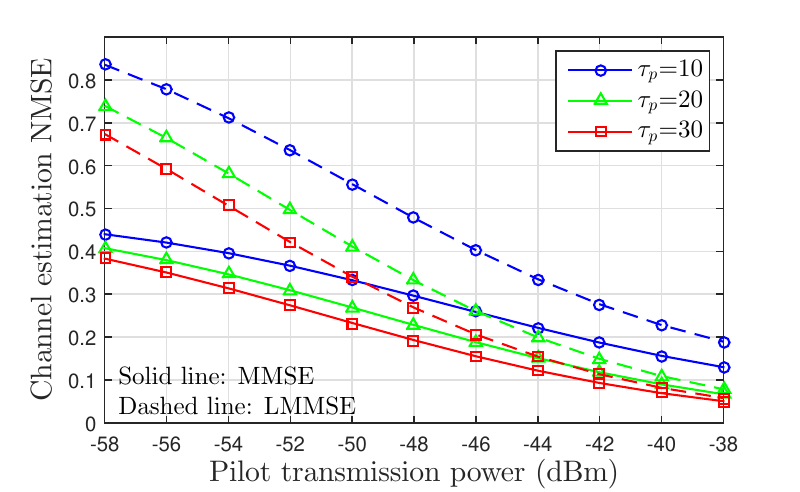}
        \caption{ }
        \label{fig:NMSE_vs_Pp_Tps}
        \end{subfigure}
        ~~
        \begin{subfigure}[b]{0.48\textwidth}
        \includegraphics[width=\textwidth]{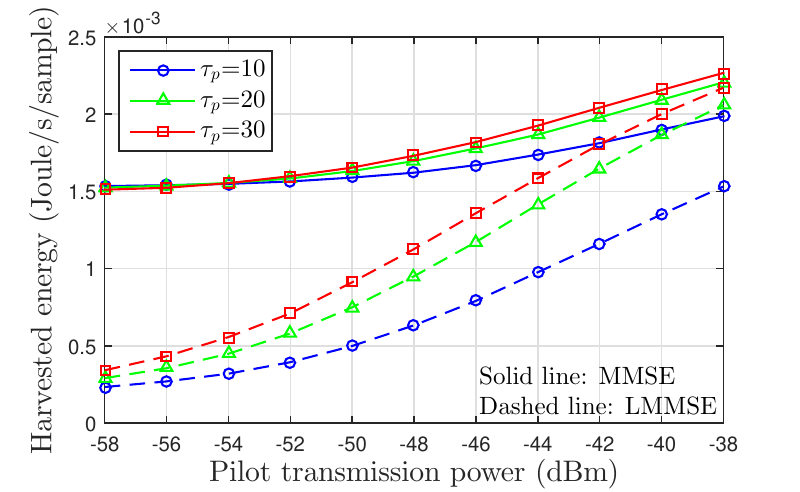}
        \caption{ }
        \label{fig:Eh_vs_Pp_Tps}
        \end{subfigure}
        ~~
        \begin{subfigure}[b]{0.48\textwidth}
        \includegraphics[width=\textwidth]{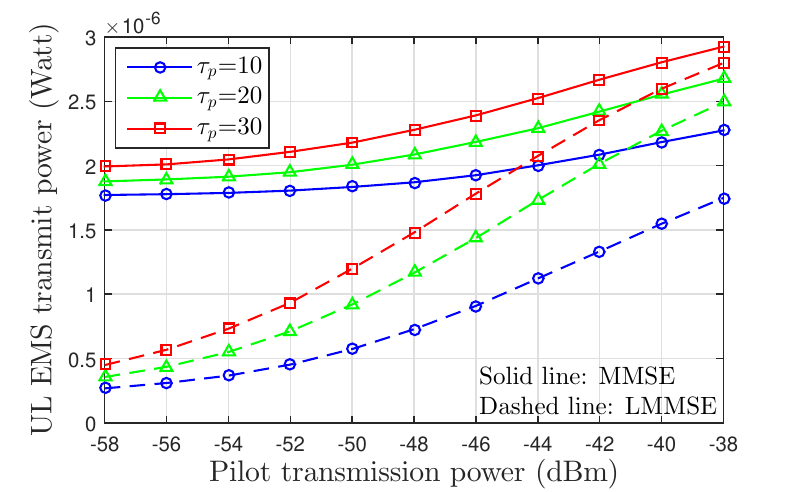}
        \caption{ }
        \label{fig:ULPw_vs_Pp_Tps}
        \end{subfigure}
        ~~
        \begin{subfigure}[b]{0.48\textwidth}
        \includegraphics[width=\textwidth]{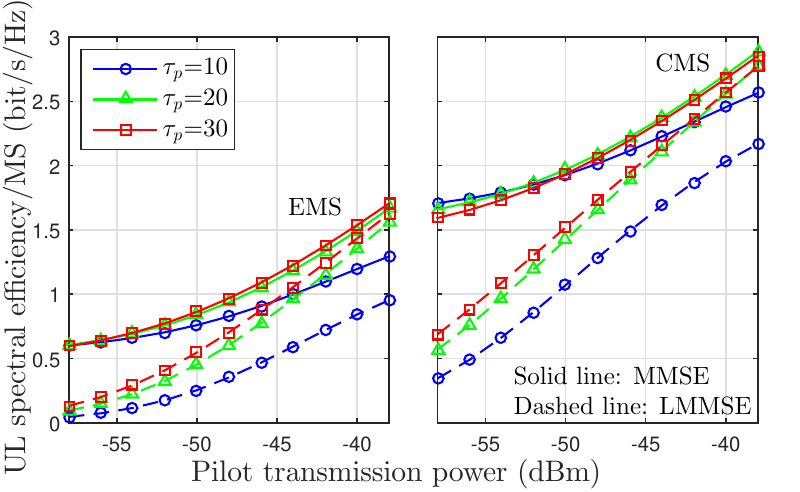}
        \caption{ }
        \label{fig:Se_vs_Pp_Tps}
        \end{subfigure}

    \caption{Average channel estimation NMSE, average harvested energy per \gls{EMS}, average transmit power per-\gls{EMS} during the phase of payload transmission, and UL spectral efficiency as a function of the pilot transmission power and using the \gls{UL} training phase length as parameter ($M=$~50 \glspl{AP}, $K_C=$~5 \glspl{CMS}, $K_E=$~10 \glspl{EMS}, $\xi_{k_E}=$~10, and $\alpha_p=$~1).}\label{fig:FvsPp_Tps}
\end{figure*}

The interplay among the pilot transmission power $P_p$, the average harvested energy, and the spectral efficiency of both the \glspl{EMS} and \glspl{CMS} is evaluated in Fig.~\ref{fig:FvsPp_Tps}. Results presented in this figure have been obtained assuming a \gls{CF-mMIMO} network with $M=$~50 \glspl{AP} serving  $K_C=$~5 \glspl{CMS} and $K_E=$~10 \glspl{EMS} with $\xi_{k_E}=$~10, and using different \gls{UL} pilot transmission phase lengths $\tau_p$ and a pilot transmit power control coefficient $\alpha_p=1$. As shown in Fig.~\ref{fig:NMSE_vs_Pp_Tps}, increasing either $P_p$ or $\tau_p$ results in a lower channel estimation error that, in turn, redounds to the implementation of higher quality beamformers. Note, moreover, that providing the system with high quality pilot signals (i.e., large values of $P_p$ and $\tau_p$) is particularly beneficial when implementing an \gls{LMMSE} channel estimation process. In fact, even though for very low values of $P_p$ and/or $\tau_p$ the \gls{MMSE} channel estimator performs much better than the \gls{LMMSE}, the performance metrics provided by both estimators become practically identical when the values of these parameters are very high. Improving the quality of the channel estimates and hence, that of the precoders during the \gls{DL} energy harvesting phase and that of the decoders during the payload transmission phase, allows for a more efficient energy harvesting process at the \glspl{EMS} and the provision of a higher spectral efficiency to both \glspl{EMS} and \glspl{CMS}, as it can be observed in Fig.~\ref{fig:Se_vs_Pp_Tps}. Given a fixed \gls{UL} pilot transmission phase length $\tau_p$, it can be clearly observed that for very low values of $P_p$ the channel estimation quality is rather poor, especially for the \gls{LMMSE} channel estimator, the amount of harvested energy at the \glspl{EMS} is modest and, as shown in Fig.~\ref{fig:ULPw_vs_Pp_Tps}, the fraction of harvested energy that these \glspl{MS} dedicate to the \gls{UL} payload data transmission phase is fairly exiguous, thus resulting in low \gls{UL} per-\gls{MS} spectral efficiencies. As $P_p$ increases, the channel estimation quality improves, the figures of harvested energy at the \glspl{EMS} are also enhanced and, even though a larger amount of harvested energy has to be allocated to transmit the pilot sequences at the \glspl{EMS}, there is still enough room to further increase the transmit power during this phase and improve the \gls{UL} spectral efficiencies. Although not shown in these figures, as $P_p$ keeps increasing, there is a certain point in which the fraction of harvested energy that is left after saving the energy necessary to transmit the pilot during the next \gls{UL} pilot transmission phase decreases and this results in a reduction in the rate of increase of the spectral efficiency that can be provided to the \glspl{EMS}. Furthermore, as the spectral efficiency that can be delivered to the \glspl{CMS} is tightly related to that supplied to the \glspl{EMS}, due to the implementation of the coupled UL/DL max-min optimization problem, increasing $P_p$ beyond the \textit{optimal} one also affects the spectral efficiency provided to the \glspl{CMS}. It is worth recalling at this point that, as previously mentioned in the system model description, the graphs presented in Fig.~\ref{fig:FvsPp_Tps} have only been plotted over the range of $P_p$ values for which it can be ensured a negligible probability that the amount of harvested energy is not sufficient to guarantee the transmission of the pilot sequence.

For a fixed value of $P_p$, increasing $\tau_p$ allows for both a better channel estimation quality and the availability of higher average harvested energy values that can be devoted to the \gls{UL} payload data transmission phase at the \glspl{EMS}. Increasing $\tau_p$, however, produces a decrease in $\tau_u$ and this affects negatively the spectral efficiency provided to the \glspl{MS}. Consequently, increasing $\tau_p$ beyond a certain limit can eventually result in a decrease of the spectral efficiency of the proposed network.

\begin{figure*}[t!]
    \centering

    \begin{subfigure}[t]{0.348\textwidth}
        \centering
        \includegraphics[height=7.2cm]{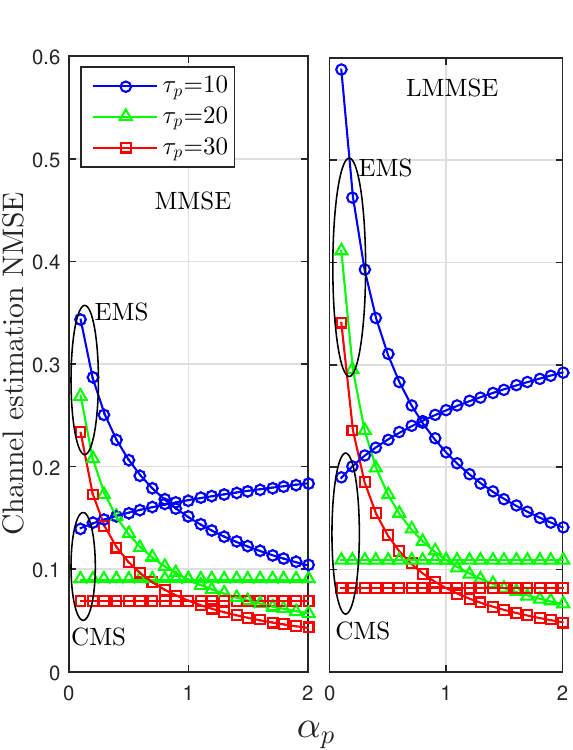}
        \caption{ }
        \label{fig:NMSE_vs_Alp_Tps}
    \end{subfigure}
    \begin{subfigure}[t]{0.318\textwidth}
        \centering
        \includegraphics[height=7.2cm]{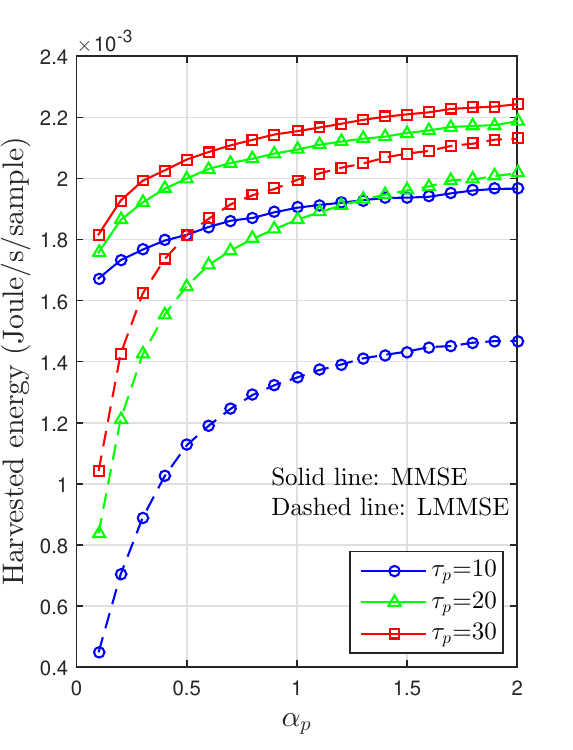}
        \caption{ }
        \label{fig:Eh_vs_Alp_Tps}
    \end{subfigure}
    \begin{subfigure}[t]{0.318\textwidth}
        \centering
        \includegraphics[height=7.2cm]{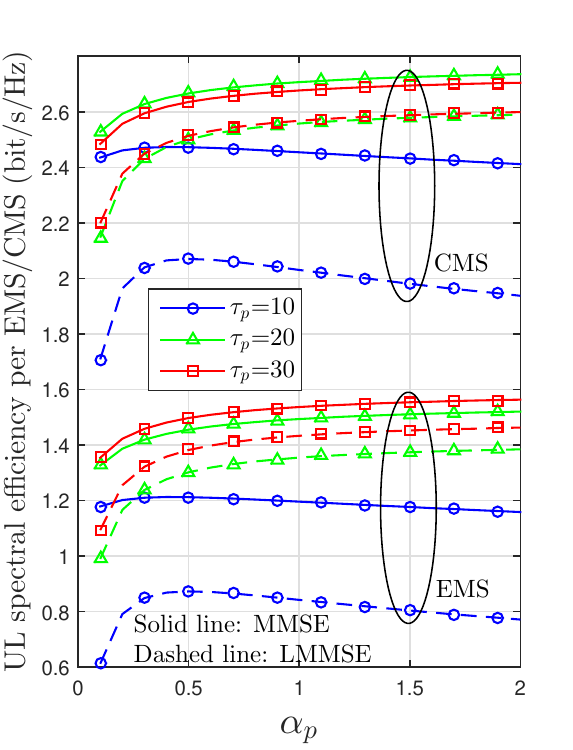}
        \caption{ }
        \label{fig:Se_vs_Alp_Tps}
    \end{subfigure}

        \caption{Average channel estimation NMSE, average harvested energy per \gls{EMS}, average transmit power per-\gls{EMS} during the phase of payload data transmission, and UL spectral efficiency as a function of the \gls{UL} power control coefficient $\alpha_p$ and using the \gls{UL} training phase length as parameter ($M=$~50 \glspl{AP}, $K_C=$~5 \glspl{CMS}, $K_E=$~10 \glspl{EMS}, $\xi_{k_E}=$~10, and $P_p=$~-40 dBm).}\label{fig:FvsAlp_Tps}
\end{figure*}

In order to further evaluate the impact the pilot transmission power has on the performance of the proposed \gls{SWIPT}-enhanced \gls{CF-mMIMO} network, the average channel estimation \gls{NMSE}, the average harvested energy at the \glspl{EMS} and the \gls{UL} spectral efficiency provided to both \glspl{EMS} and \glspl{CMS} are depicted in Fig.~\ref{fig:FvsAlp_Tps} as a function of the \gls{UL} power control coefficient $\alpha_p$ and using the \gls{UL} training phase length as parameter. Results presented in this figure have been obtained assuming a \gls{CF-mMIMO} network with $M=$~50 \glspl{AP} serving $K_C=$~5 \glspl{CMS} and $K_E=$~10 \glspl{EMS} with $\xi_{k_E}=$~10, and a pilot transmit power $P_p=$~-40 dBm. First of all, notice that the channel estimation \gls{NMSE} previosly represented in Figs.~\ref{fig:MMSE_vs_Tp} and \ref{fig:NMSE_vs_Pp_Tps} was obtained assuming an \gls{UL} power control coefficient $\alpha_p=$~1. That is, both \glspl{EMS} and \glspl{CMS} were assumed to use exactly the same pilot transmission power and, hence, both groups of \glspl{MS} were characterized with exactly the same average channel estimation quality. Using a value of $\alpha_p\neq$~1, in contrast, splits the \glspl{MS} in two groups experiencing dissimilar qualities in the channel estimation process. Indeed, as shown in Fig.~\ref{fig:NMSE_vs_Alp_Tps}, increasing $\alpha_p$ results in a substantial improvement of the average channel estimation \gls{NMSE} for the \glspl{EMS} but, in those cases where $\tau_p < K$ and there is pilot contamination, it reverts into a slight decrease in the average channel estimation \gls{NMSE} experienced by the \glspl{CMS} for both \gls{MMSE} and \gls{LMMSE} channel estimators. Again, it is clearly shown in this figure that increasing the length of the pilot sequences improves the quality of the channel estimates irrespective of the group of \glspl{MS} under evaluation. Now, turning our attention to results presented in Figs.~\ref{fig:Eh_vs_Alp_Tps} and \ref{fig:Se_vs_Alp_Tps} for a fixed \gls{UL} pilot transmission phase length $\tau_p$, it can be observed that \glspl{EMS} are characterized with a rather poor channel estimation quality when using low $\alpha_p$ values. This results in a modest amount of harvested energy at the \glspl{EMS} and, due to the coupled \gls{UL}/\gls{DL} max-min optimization problem, it also reverts onto fairly low spectral efficiency metrics at both \glspl{EMS} and \glspl{CMS}. The channel estimation quality and the quantity of harvested energy at the \glspl{EMS} improve as $\alpha_p$ increases and, again, even though a larger amount of harvested energy has to be allocated to transmit the pilot sequences at the \glspl{EMS} and the channel estimation quality slightly decreases at the \glspl{CMS}, there is still room for \gls{UL} spectral efficiency improvement. As $\alpha_p$ keeps increasing, most of the harvested energy at the \glspl{EMS} has to be allocated to the pilot transmission during the \gls{UL} training phase and the fraction of harvested energy that can be allocated to transmit the \gls{UL} payload data gets smaller and smaller hence forcing a diminishing rate of increase (or even a decrease) of the \gls{UL} spectral efficiencies provided to both groups of \glspl{MS}, especially notable for low values of $\tau_p$.

Again, for a fixed value of $\alpha_p$, a better channel estimation quality and a higher average harvested energy can be provided when increasing $\tau_p$. Increasing $\tau_p$, however, is done at the cost of decreasing $\tau_u$ and this may eventually result in a reduction of the \gls{UL} spectral efficiency provided to both \glspl{EMS} and \glspl{CMS}.

\subsection{Impact of \gls{MS} weighting coefficients on system energy/spectral efficiencies}

\begin{figure*}[t!]

    \centering
    \begin{subfigure}[t]{0.328\textwidth}
        \centering
        \includegraphics[height=7.2cm]{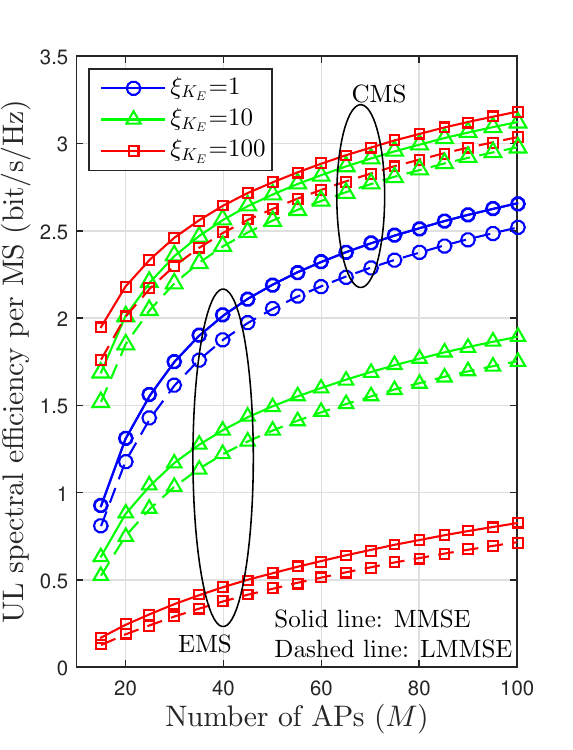}
        \caption{ }
        \label{fig:SepMS_vs_M_xiEs}
    \end{subfigure}
    \begin{subfigure}[t]{0.328\textwidth}
        \centering
        \includegraphics[height=7.2cm]{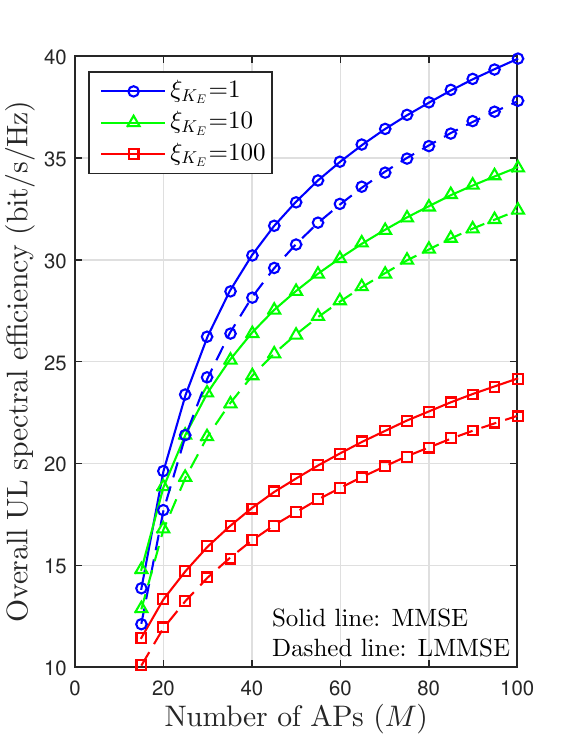}
        \caption{ }
        \label{fig:OSe_vs_M_xiEs}
    \end{subfigure}
    \begin{subfigure}[t]{0.328\textwidth}
        \centering
        \includegraphics[height=7.2cm]{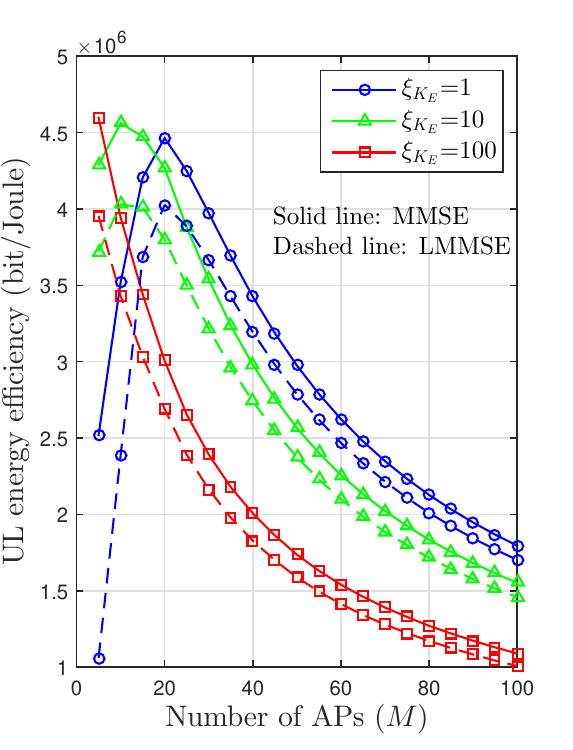}
        \caption{ }
        \label{fig:OEe_vs_M_xiEs}
    \end{subfigure}

        ~~

    \caption{\Gls{UL} spectral efficiency per \gls{MS}, overall \gls{UL} spectral efficiency, and \gls{UL} energy efficiency as a function of the number of \glspl{AP} in the network and with the weighting coefficient factor $\xi_{k_E}$ as parameter ($K_C=$~5 \glspl{CMS}, $K_E=$~10 \glspl{EMS}, $P_p=$~-40 dBm, and $\alpha_p=$~1).}
    \label{fig:FvsM_KCs_xiEs}
\end{figure*}

Finally, the impact on the \gls{UL} spectral and energy efficiencies produced by the use of different weighting coefficient factors $\xi_{k_E}$ in the coupled \gls{UL}/\gls{DL} max-min optimization problem is evaluated in Fig.~\ref{fig:FvsM_KCs_xiEs} as a function of the number of \glspl{AP} in the network. Results presented in this figure have been obtained assuming a \gls{SWIPT}-enhanced \gls{CF-mMIMO} network serving $K_C=$~5 \glspl{CMS} and $K_E=$~10 \glspl{EMS} with $\xi_{k_E}=$~10, a pilot transmit power $P_p=$~-40 dBm and a pilot power control coefficient $\alpha_p=$~1. As shown in Fig.~\ref{fig:SepMS_vs_M_xiEs}, using a weighting coefficient factor $\xi_{k_E}=$~1, the network provides exactly the same spectral efficiency to both \glspl{EMS} and \glspl{CMS}. Increasing the weighting coefficient factor beyond $\xi_{k_E}=$~1 prioritizes the achievable \gls{SINR} values provided to the \glspl{CMS} over those provided to the \glspl{EMS} and, in addition, given the logarithmic relationship between the spectral efficiency and the achievable \gls{SINR}, the relative reduction in spectral efficiency experienced by the \glspl{EMS} is much more significant than the corresponding increment experienced by \glspl{CMS}, especially for scenarios with a high spatial density of \glspl{AP}. The combination of a modest relative increase in spectral efficiency experienced by the \glspl{CMS} and a dramatic decrease in spectral efficiency experienced by the \glspl{EMS} results in the overall spectral efficiency shapes represented in Fig.~\ref{fig:OSe_vs_M_xiEs}. Notably, a dramatic drop in overall spectral efficiency can be expected when increasing $\xi_{k_E}$ irrespective of the density of \glspl{AP} per area unit. The most interesting effects produced by the modification of $\xi_{k_E}$, however, can be observed in Fig.~\ref{fig:OEe_vs_M_xiEs} representing the \gls{UL} energy efficiency as a function of the number of \glspl{AP} in the network. Increasing the number of serving \glspl{AP}, as shown in Fig.~\ref{fig:OSe_vs_M_xiEs}, helps improving the overall spectral efficiency but, as has been shown in Subsection \ref{subsec:ULPowCons}, it also produces a considerable increase in the \gls{UL} power consumption. As the rate of increase in power consumption is almost linear with $M$ and the rate of increase in overall spectral efficiency is much modest, the combined effects result in the \gls{UL} energy efficiency behaviour represented in Fig.~\ref{fig:OEe_vs_M_xiEs}. Note how the \gls{UL} achievable energy efficiency increases when reducing the spatial density of \glspl{AP} up to a certain value where an optimum is reached which, for this particular scenario, is located around $M=$~4, 10 and 20 \glspl{AP} for $\xi_{k_E}=$~100, 10 and 1, respectively. Selecting an appropriate value of $\xi_{k_E}$ will obviously depend on the particular setup of the scenario under evaluation and on which can be considered a good trade-off among sometimes conflicting \gls{QoS} metrics including, among others, the spectral efficiency, the energy efficiency, or the amount of harvested energy at the \glspl{EMS}.

\section{Conclusion}
\label{sec:Conclusion}

A \gls{SWIPT}-enhanced \gls{CF-mMIMO} network has been proposed where a large number of spatially distributed multi-antenna \glspl{AP} interconnected via a CPU collaborate in serving a large number of single-antenna \glspl{EMS} (requiring wireless energy transfer) and \glspl{CMS} (not requiring wireless energy transfer) on the same time-frequency resources. Spatially correlated Rician fading channels and \gls{ZF} processing based on either \gls{MMSE} or \gls{LMMSE} channel estimation have been considered. Mathematically manageable expressions have been derived for the harvested energy during the \gls{DL} energy harvesting phase and the \textit{achievable} spectral and energy efficiencies during the \gls{UL} payload transmission phase. Moreover, a coupled \gls{UL}/\gls{DL} optimization problem has been formulated that aims at finding the power control coefficients maximizing the minimum of the weighted \gls{UL} achievable \glspl{SINR} of all \glspl{MS}. Extensive numerical results have revealed the existing trade-offs among the achievable spectral and energy efficiencies, the amount of harvested energy at the \glspl{EMS}, the quantity of energy dedicated to \gls{UL} pilot transmission or the network setup under consideration. In particular, although the average harvested energy per \gls{EMS} increases with both the training phase length $\tau_p$ and the \gls{DL} energy harvesting phase length $\tau_h$, there are optimal values of both $\tau_p$ and $\tau_h$ that increase with the number of \glspl{EMS} in the network and decrease with the number of serving \glspl{AP}. Furthermore, the optimal values can be very different depending on whether the objective function is the aggregate spectral efficiency or the spectral efficiency experienced by a particular group of \glspl{MS} (i.e., \glspl{EMS} or \glspl{CMS}). A trade-off among the pilot transmit power $P_p$, the pilot power control coefficient $\alpha_p$ and the performance metrics of the network has also been uncovered. Increasing $P_p$ helps improving the quality of both the channel estimates and the beamformers, however, if this is increased beyond reasonable values, most of the harvested energy at the \glspl{EMS} has to be allocated to the pilot transmission during the \gls{UL} training phase and the smaller fraction of harvested energy that can be allocated to transmit the \gls{UL} payload data forces a diminishing rate of increase (or even a decrease) of the \gls{UL} spectral efficiencies provided to the \glspl{MS}. Similar effects can be expected when increasing $\alpha_p$, except that in this case the quality of the channel estimates and beamformers is only improved for the \glspl{EMS}. Moreover, it has been shown that, under the maximum energy efficiency criterion, there is an optimal number of serving \glspl{AP} that also depends on the particular scenario under evaluation and increases with the weighting coefficient factor $\xi_{k_E}$.

There are some open issues related to the practical implementation of the proposed \gls{SWIPT}-enhanced \gls{CF-mMIMO} network that remain to be addressed in future works. In particular, it could be very interesting to consider the use of limited-capacity fronthaul links between the \glspl{AP} and the \gls{CPU}, the design of joint \gls{AP} clustering and \gls{MS} scheduling strategies, the optimization of energy efficiency-related optimization problems, or the design of radio resource management strategies to cope with those situations in which the energy available at a given \gls{EMS} is not enough to warrant the transmission of either data symbols and/or pilot sequences.

\bibliographystyle{IEEEtran}
\bibliography{Cell_free_biblio}

\end{document}